\documentclass[amsmath,amssymb,12pt,reprint,nofootinbib, onecolumn,longbibliography]{revtex4-1}

\usepackage{xcolor}
\usepackage{mathtools}
\usepackage{comment}
\usepackage{caption}
\usepackage{subcaption}
\usepackage{placeins}

\usepackage{amsmath}
\usepackage{amsfonts}
\usepackage{mathrsfs}
\usepackage{chemformula}

\usepackage{makecell}

\usepackage{soul} 

\newcommand{\bs}{\boldsymbol}

\newcommand{\kb}{k_{\text{B}}}

\newcommand{\position}{\boldsymbol{x}}
\newcommand{\absposition}{x}
\newcommand{\paramposition}{\tilde{x}}
\newcommand{\gradposition}{\boldsymbol{\nabla}_x}

\newcommand{\spatial}{\boldsymbol{r}}
\newcommand{\absspatial}{r}
\newcommand{\gradspatial}{\boldsymbol{\nabla}_r}
\newcommand{\laplacian}{\boldsymbol{\nabla}_r^2}

\newcommand{\graddiscrete}{\boldsymbol{\nabla}_{h}}

\newcommand{\force}{\boldsymbol{F}}
\newcommand{\forcedensity}{\boldsymbol{f}}
\newcommand{\commentout}[1]{}

\def\shrug{\texttt{\raisebox{0.75em}{\char`\_}\char`\\\char`\_\kern-0.5ex(\kern-0.25ex\raisebox{0.25ex}{\rotatebox{45}{\raisebox{-.75ex}"\kern-1.5ex\rotatebox{-90})}}\kern-0.5ex)\kern-0.5ex\char`\_/\raisebox{0.75em}{\char`\_}}}

\newcommand{\cutoff}{\psi}

\newcommand{\particlenoise}{\boldsymbol{\mathcal{W}}}
\newcommand{\discreteparticlenoise}{\boldsymbol{W}}

\newcommand{\mobility}{\boldsymbol{\mathcal{M}}}

\newcommand{\radius}{a}

\newcommand{\fluidvel}{\boldsymbol{v}}
\newcommand{\particlevel}{\boldsymbol{V}}

\newcommand{\fluidvelabs}{v}

\newcommand{\fieldnoise}{\boldsymbol{\mathcal{Z}}}
\newcommand{\discretefieldnoise}{\boldsymbol{Z}}
\newcommand{\fulldiscretenoise}{\widehat{\boldsymbol{Z}}}

\newcommand{\hydrospread}{\bs{\mathcal{S}}^\mathrm{hy}}
\newcommand{\hydrointerp}{\bs{\mathcal{J}}^\mathrm{hy}}

\newcommand{\electrospread}{\bs{\mathcal{S}}^\mathrm{es}}
\newcommand{\electrointerp}{\bs{\mathcal{J}}^\mathrm{es}}

\newcommand{\stokes}{\bs{\mathcal{L}}}

\newcommand{\efield}{\bs{E}}

\newcommand{\chargevec}{\boldsymbol{q}}
\newcommand{\charge}{q}

\newcommand{\hydrokernel}{\delta^\mathrm{hy}}
\newcommand{\electrokernel}{\delta^\mathrm{es}}

\newcommand{\chargedensity}{\varrho}

\newcommand{\wallnormal}{\hat{\boldsymbol{n}}}
\newcommand{\walltangent}{\hat{\boldsymbol{\tau}}}

\newcommand{\cellvol}{\Delta \mathcal{V}}

\newcommand{\ito}{ }

\newcommand{\Dlength}{\lambda_\text{D}}
\newcommand{\GClength}{\lambda_\text{GC}}

\newcommand{\NaPos}{{\mathrm{Na}^+}}
\newcommand{\ClNeg}{{\mathrm{Cl}^-}}

\newcommand{\DISCOS}{DISCOS}

\newcommand{\deleted}[1]{}

\newcommand{\MarginPar}[1]   
{\marginpar{\vskip-\baselineskip 
\raggedright\tiny\sffamily\hrule\smallskip{\color{red}#1}\par\smallskip\hrule}}

\newcommand{\MarginJBB}[1]   
{\marginpar{\vskip-\baselineskip 
\raggedright\tiny\sffamily\hrule\smallskip{\color{blue}#1}\par\smallskip\hrule}}

\begin{document}

\title[Electrokinetic flows with \DISCOS]
      {Modeling Electrokinetic Flows with the Discrete Ion Stochastic Continuum Overdamped Solvent
      Algorithm}

\author{D. R. Ladiges}
\email{DRLadiges@lbl.gov}
\affiliation{Center for Computational Sciences and Engineering, LBNL}
\homepage{https://ccse.lbl.gov}

\author{J. G. Wang}
\affiliation{Center for Computational Sciences and Engineering, LBNL}

\author{I. Srivastava }
\affiliation{Center for Computational Sciences and Engineering, LBNL}

\author{S. P. Carney}
\affiliation{Department of Mathematics, UCLA}

\author{A. Nonaka}
\affiliation{Center for Computational Sciences and Engineering, LBNL}

\author{A. L. Garcia}
\affiliation{Department of Physics and Astronomy, SJSU}

\author{A. Donev}
\affiliation{Courant Institute of Mathematical Sciences, New York University, New York, NY 10012}

\author{J. B. Bell}
\affiliation{Center for Computational Sciences and Engineering, LBNL}

\date{\today}

\begin{abstract}
In this article we develop an algorithm for the efficient simulation of electrolytes in the presence of physical boundaries. In previous work the Discrete Ion Stochastic Continuum Overdamped Solvent (DISCOS) algorithm was derived for triply periodic domains, and was validated through ion-ion pair correlation functions and Debye-H{\"u}ckel-Onsager theory for conductivity, including the Wien effect for strong electric fields.
In extending this approach to include an accurate treatment of physical boundaries we must address several important issues. First, the modifications to the spreading and interpolation operators necessary to incorporate interactions of the ions with the boundary are described. Next we discuss the modifications to the electrostatic solver to handle the influence of charges near either a fixed potential or dielectric boundary. An additional short-ranged potential is also introduced to represent interaction of the ions with a solid wall. Finally, the dry diffusion term is modified to account for the reduced mobility of ions near a boundary, which introduces an additional stochastic drift correction.
Several validation tests are presented confirming the correct equilibrium distribution of ions in a channel. Additionally, the methodology is demonstrated using electro-osmosis and induced charge electro-osmosis, with comparison made to theory and other numerical methods. Notably, the DISCOS approach achieves greater accuracy than a continuum electrostatic simulation method. We also examine the effect of under-resolving hydrodynamic effects using a `dry diffusion' approach, and find that considerable computational speedup can be achieved with a negligible impact on accuracy.
\end{abstract}

\maketitle

\section{Introduction}\label{sec:Introduction}

The ability to model electrolytes in confined systems is fundamental to the understanding of processes such as electrophoresis, electro-osmosis, and electrochemistry. These arise in a wide range of circumstances, including biological systems \cite{grodzinsky2011fields} and engineered devices \cite{kirby2010micro,hernandez2022thermodiffusive}
such as catalytic micropumps \cite{verpoorte2002microfluidic,Li2019}, batteries \cite{bachman2016inorganic,Scott2018}, and fuel cells \cite{Andersson_2010,JahnkeETAL_2016,Arsalis2019}. Many of these phenomena occur at scales where thermal fluctuations play a significant role, and the need to capture these fluctuations has spurred the development of a range of different numerical methods for mesoscale modeling of electrolytes. 
These include purely continuum methods \cite{peraud2016low,donev2019fluctuating,donev2019fluctuating2} based on generalizations of fluctuating hydrodynamics (FHD) \cite{Land2}, as well as molecular dynamics models that feature either an implicit \cite{onufriev2008implicit} or explicit \cite{frenkel2001understanding} treatment of the solvent molecules. Other methods combine these approaches by employing a discrete description of particles in a fluctuating, continuum solvent. Some examples include the General Geometry Ewald-like Method (GGEM) 
 \cite{hernandez2007fast,zhao2017parallel}, the Stochastic Eulerian Lagrangian Method (SELM) \cite{atzberger2011, atzberger2007stochastic}, the Stochastic Force Coupling Method (SFCM) \cite{maxey2001localized, lomholt2003force, keaveny2014fluctuating, delmotte2015simulating}, and the Fluctuating Immersed Boundary (FIB) method \cite{delong2014brownian}. Some discussion of the relationship between these methods is given in Refs.~\citenum{delong2014brownian} and \citenum{ladiges2020discrete}. While GGEM employs Green's functions in conjunction with a Stokes solver to apply local corrections, the remaining methods apply some form of particle-grid coupling to maintain a Lagrangian description of the particles and an Eulerian description of the solute solvent.
In this vein, the FIB method uses Immersed Boundary (IB) \cite{peskin2002} kernels to couple particles to an explicitly simulated fluctuating solvent, enabling Brownian dynamics (BD) simulations without the use of Green's functions. In doing so many of the difficulties associated with traditional BD methods, such as the need to construct and invert the mobility matrix, are avoided. 

More recently, we proposed an extension of the FIB approach called the Discrete Ion Stochastic Continuum Overdamped Solvent (DISCOS) method \cite{ladiges2020discrete}. This approach models electrolytes by employing a discrete description of the ions, while modeling the solvent using the overdamped fluctuating Stokes equations. In the overdamped limit, ion motion is computed by spreading the forces on the particles to a grid via the IB method, solving Stokes equation to compute the velocity field induced by those
forces, and then interpolating those velocities back onto the particles. 
The FIB approach is
extended through the use of `dry diffusion' \cite{espanol2015coupling}, which yields improved computational speed by allowing some under-resolution of the hydrodynamic grid without significant loss of accuracy.
An immersed boundary (IB) variant of the particle-particle, particle-mesh (P3M) \cite{Hockney:1988:CSU:62815, frenkel2001understanding} approach is used to calculate electrostatic interactions, and short range particle interactions are calculated directly using an interaction potential such as  Weeks-Chandler-Andersen (WCA) \cite{Weeks1971}. In Ref.~\citenum{ladiges2020discrete} this approach was described for triply periodic domains, and shown to accurately reproduce ion-ion pair correlation functions, and Debye-H{\"u}ckel-Onsager theory for conductivity, including the Wien effect for strong electric fields.

\textcolor{black}{Although a range of properties of bulk electrolytes can be studied using triply-periodic domains, as discussed above, many physical systems of interest feature confining boundaries. Aside from those mentioned above, this also includes metal electroplating \cite{PhysRevE.48.1279}, and systems utilizing electrodialysis \cite{urtenov2013basic,soniat2021toward} such as desalinization \cite{nikonenko2014desalination} and the utilization of solar energy to produce fuels \cite{zhang2011electrochemical,fabian2015particle}.} In this article we extend the DISCOS method to include flows in the presence of confining boundaries, including slip and no-slip walls, and where each location on a wall is held at a specified electric potential or treated as weakly polarizable dielectric with a specified surface charge density.

This requires a number of non-trivial modifications to the approach described in Ref.~\cite{ladiges2020discrete}. First, the finite volume discretization of the Stokes and Poisson solvers must be modified to treat Dirichlet or Neumann boundaries. In particular, care must be taken to ensure the discretization near the boundary satisfies a discrete version of the fluctuation-dissipation theorem, as discussed in Ref.~\cite{balboa2012staggered}. 
Second, the method of images is employed when a particle gets sufficiently close to a physical boundary to correct for the fact that the particle's immersed boundary kernel will overlap with the boundary. 
This occurs both when computing an ions' hydrodynamic interaction with a solid boundary and when calculating the particle-mesh portion of the P3M method. 
In the presence of boundaries when the hydrodynamics is underresolved, the dry diffusion becomes
a `dry' diffusion tensor to reflect the anisotropic changes in particle mobility as the particle nears the boundary. 
The dependence of the dry diffusion tensor on particle positions leads to an additional It\^{o} drift that is needed to ensure the system achieves detailed balance. 
Finally, an additional short range potential is included to represent particle interactions with the boundaries.

The layout of this paper is as follows. First in Section~\ref{sec:DISCOS} we describe the DISCOS approach
with an emphasis on the modifications needed for including physical boundaries.
In Section~\ref{sec:Results} we perform some numerical validation of DISCOS by comparison to theory and simulations that appear in the literature. This includes a test of the methods we have used to incorporate the stochastic drift correction. In Sec.~\ref{sec:electrokinetic} we demonstrate the application of DISCOS to electrokinetic flows. First we examine electro-osmosis in a channel, with comparison to theory, a deterministic continuum numerical method, and molecular dynamics (MD) approach. Also included is an analysis of the use of dry diffusion. This is followed by a demonstration of the use of DISCOS to compute the complex flow patterns of induced charge electro-osmosis (ICEO). Finally, Section~\ref{sec:conclusion} contains some discussion and concluding remarks.

\section{The DISCOS method}
\label{sec:DISCOS}

This section describes the DISCOS method, with an emphasis on the changes required for incorporating boundary conditions.
For a more detailed discussion of the periodic case see Ref.~\citenum{ladiges2020discrete}.
First, in Section~\ref{sec:BD} we describe the overall 
fluctuating hydrodynamic framework followed by 
the computation of forces on the particles in Section~\ref{subsec:Electrostatics}.
In Section~\ref{sec:drymob} we outline the incorporation of ``dry diffusion", an approach that adds additional diffusion to the particles relative to the coarse-grained velocity as discussed in Ref.~\citenum{espanol2015coupling}. Finally in Section~\ref{subsec:TemporalAlgorithm} we combine each of these parts into the overall DISCOS algorithm.

\subsection{Brownian dynamics}\label{sec:BD}
 
Fort $N$ ions with positions $\position(t) = \{\position_1(t),\cdots,\position_i(t),\cdots,\position_N(t)\}$
the Brownian dynamics equation of motion, expressed in It\^o form, is
\cite{klimontovich1990ito,hutter1998fluctuation}
\begin{equation}
\frac{d\position}{dt} =\mobility \force    + \kb T \gradposition \cdot \mobility   + \sqrt{2 \kb T} \mobility^{1/2}\ito \particlenoise,
\label{eq:brownian2}
\end{equation}
\textcolor{black}{where $\force(\position) = \left\{\force_1(\position_1),\cdots,\force_i(\position_i),\cdots,\force_N(\position_N)\right\}$ are the forces on the particles, $\kb$ is Boltzmann's constant, $T$ is temperature, $\gradposition$ is the gradient operator with respect to particle positions, and $\particlenoise(t) =  \left\{\particlenoise(t)_1,\cdots,\particlenoise(t)_i,\cdots,\particlenoise(t)_M\right\}$}
\footnote{Note that $M$ need not equal $N$, since the number of noise terms $M$ corresponds to the number of hydrodynamic grid points, not the number of particles. The mobility matrix $\mobility= \mobility^{1/2}(\mobility^{1/2})^\star$ is a square matrix of size $3N\times 3N$, however, $\mobility^{1/2}$ and its transpose-conjugate $(\mobility^{1/2})^\star$ are generally nonsquare. }
are standard independent Gaussian white noise processes on $\mathbb{R}^3$.  
The symmetric positive-definite mobility matrix $\mobility(\position)$ encodes the hydrodynamic interactions between particles.
The divergence of the mobility term on the right hand side of Eq.~(\ref{eq:brownian2}) is a stochastic drift term
that arises from writing the system in It\^o form.

Similarly to the FIB algorithm discussed in Ref.~\citenum{delong2014brownian}, DISCOS computes the hydrodynamic interactions
between particles using a finite volume fluctuating hydrodynamics solver instead of Green's functions.  In this approach,
forces computed on the particles are \emph{spread} to the finite volume grid, the hydrodynamic equations are solved to compute
the velocity field induced by those forces, and, finally, the velocity is \emph{interpolated} back onto the particles so
their positions can be updated.

For the mapping between the Lagrangian particles and the Eulerian fields we employ the immersed boundary (IB) method \cite{peskin2002}. 
The spatial extent of each particle is defined by a compactly supported kernel function, $\hydrokernel(\spatial)$; see Appendix \ref{appdx2} for several examples.
Spreading and interpolation operators are then defined in terms of these IB kernels.  

The discrete spreading operator, $\hydrospread_h$, maps the particle forces, $\force$, onto the grid such that
\begin{align}
\forcedensity^\text{s}(\spatial_j) =&\, \left(\hydrospread_h\force\right)(\spatial_j) = \sum_{i=1}^N 
\hydrokernel(\position_i - \spatial_j)\force_{i},\label{kernel2}
\end{align}
where the vector $\forcedensity^\text{s}(\spatial_j)$ represents the spread force density at discrete grid location $\spatial_j$. The interpolation operator, $\hydrointerp_h$, maps the fluid velocity on the grid, $\fluidvel$, to particle locations as
\begin{align}
\particlevel_i =&\,
\left(\hydrointerp_h \bs{v}\right)_i  =\Delta\mathcal{V}\sum_{j\in\Omega_i^\mathrm{Pe}}  \hydrokernel(\position_i - \spatial_j)\fluidvel(\spatial_j).\label{kernel1}
\end{align}
Here, $\cellvol$ is the (constant) computational cell volume, 
$\particlevel = \left\{\particlevel_1,\cdots,\particlevel_i\cdots \particlevel_N\right\}$
denotes the velocity of the fluid interpolated to particle locations $\position$, and $\Omega_i^\mathrm{Pe}$ corresponds to the set of mesh locations in the support of $\hydrokernel$ centered at
$\position_i$.
Note that these operators satisfy an adjoint-like condition
\begin{equation}
\left( \hydrointerp_h\bs{v}\right) \cdot \force = 
 \Delta\mathcal{V} \left( \hydrospread_h\force\right) \cdot \fluidvel,
 \label{eq:adjoint}
\end{equation}
which is required for the method to conserve energy and satisfy fluctuation dissipation balance.

DISCOS computes the hydrodynamic response to 
a force on the immersed ion in the limit of asymptotically large Schmidt number,
$
\mathrm{Sc} = {\displaystyle \eta}/{\displaystyle \rho D} \gg 1 
$;
here $D$ is the diffusion coefficient of a particle, and $\eta$ and $\rho$ are the viscosity and density of the solvent, respectively.  In this  regime, the diffusion time of the particles is large compared to the relaxation time of the fluid, so the flow can be treated as quasi-steady.
The solvent, which is also assumed to be  isothermal and incompressible, can therefore
be modeled by the fluctuating Stokes equations\footnote{\textcolor{black}{Note that an
It\^o drift term like that in Eq.~(\ref{eq:brownian2}) arises after eliminating the momentum; this is not included in Eq.~(\ref{eq:stokes}).}}
\begin{subequations}
\begin{align}
\bs \nabla_r p - \eta \laplacian \fluidvel =&\, \forcedensity + \sqrt{2 \kb T \eta}\, \gradspatial \cdot \fieldnoise,\label{eq:stokesA}\\
\bs{\nabla}_r \cdot \fluidvel =&\, 0,
\end{align}\label{eq:stokes}%
\end{subequations}%
where $\bs \nabla_r$ is the gradient operator with respect to the spatial variable $\spatial$, 
$p(\spatial)$ is the pressure, and 
$\forcedensity(\spatial)$ is a force density applied to the fluid (which includes $\forcedensity^\text{s}$). Finally, $\fieldnoise(\spatial)$ is a symmetric Gaussian white noise tensor field.

DISCOS uses a second-order finite volume discretization of the Stokes equation, Eq.~(\ref{eq:stokes}), 
\begin{subequations}
\begin{align}
\graddiscrete p - \eta \graddiscrete^2 \fluidvel =&\, \forcedensity+ \sqrt{\frac{2 \kb T \eta}{\cellvol }}\, \graddiscrete\cdot \discretefieldnoise,\label{stokes_mollified}\\
\graddiscrete \cdot  \bs v =&\, 0,
\end{align}\label{eq:disc_stokes}%
\end{subequations}
where the subscript $h$ is used to indicate the discrete form of an operator.
Here, the term $\discretefieldnoise(t)$ is a finite dimensional collection of white noise processes representing the spatial discretization of $\fieldnoise$ on a regular grid with positions $\spatial_j$ and spacing $\Delta r$, with the factor $1/\sqrt{\cellvol}$ arising from the spatial coarse graining of
noise onto the mesh.
We use a staggered grid system
with normal velocities and the associated force densities defined at cell faces, and pressure defined in cell centers, i.e., a standard marker-and-cell (MAC) discretization \cite{Cai2014}.
The form of these discretizations are designed to preserve fluctuation dissipation balance, as discussed in Refs.~\citenum{delong2014brownian} and \citenum{ladiges2020discrete}.  

For problems with confining boundaries we need to specify boundary conditions for the Stokes equations.
For a no-slip hydrodynamic boundary, we apply a Dirichlet condition to all components of the velocity. For example, for a stationary wall 
$\fluidvel (\spatial_b)=0$,
where $\bs{r}_b$ is the coordinate of a point on the boundary. 
For a slip boundary we apply this condition only to the normal component, and a zero stress Neumann condition to the parallel components, giving
\begin{align}
%
 \fluidvelabs(\spatial_b)\cdot\wallnormal=0, \quad \walltangent \cdot \left.\gradspatial \fluidvelabs(\spatial)\right|_{\spatial=\spatial_b}\cdot \wallnormal =0,
\label{fluidbc}
\end{align}
where $\wallnormal$ is \textcolor{black}{the unit inward normal vector for the boundary}, and $\walltangent$ is any unit tangent vector to the boundary. Applying these boundary conditions to the solution of Eq.~(\ref{eq:disc_stokes}) is straightforward and is discussed in the context of FHD in Ref.~\citenum{balboa2012staggered}.

We next define the discrete Stokes operator $\stokes_h$ with appropriate boundary conditions such that the solution of
Eq.~\eqref{eq:disc_stokes} is
\begin{equation}
\fluidvel = \stokes^{-1}_h \left(\forcedensity^{\rm s}+\forcedensity^{\rm th} + \sqrt{\frac{2 \kb T \eta}{\cellvol }}\, \graddiscrete\cdot   \discretefieldnoise \right),
\end{equation}
where we have taken the forcing term $\forcedensity$ to consist of the spread force and an additional `thermal forcing' term which is described below. The particles' hydrodynamic velocities in terms of the particle forces is then
\begin{align}
\particlevel =& \hydrointerp_h\stokes_h^{-1}\left(\hydrospread_h\force+\forcedensity^{\rm th} + \sqrt{\frac{2 \kb T \eta}{\cellvol }}\, \graddiscrete\cdot \ito \discretefieldnoise\right).\label{eq:ions2}
\end{align}
We can now draw a correspondence with the first and third terms on the right hand side of
Eq.~(\ref{eq:brownian2}) with
\begin{equation}
 \mobility \equiv  \hydrointerp_h\stokes^{-1}_h \hydrospread_h,
 \label{eq:mob_SHIBA}
\end{equation}
and
\begin{align}
 \mobility^{1/2} \particlenoise \equiv \sqrt{\frac{2 \kb T \eta}{\cellvol }} \hydrointerp_h\stokes^{-1}_h  \graddiscrete\cdot\discretefieldnoise.
 \label{eq:sqrtmob_SHIBA}
\end{align}
The second term on the right hand side of Eq.~\eqref{eq:brownian2}, the stochastic drift term, can be decomposed as
\begin{align}
 \left(\kb T\right) \gradposition \cdot \mobility =  \left(\kb T \right) \hydrointerp_h\stokes^{-1}_h \gradposition\cdot \hydrospread_h+ \left( \kb T\right)  \left( \gradposition \hydrointerp_h\right):\left(\stokes^{-1}_h \hydrospread_h\right).\label{eq:difbypart}
\end{align}
The first term on the right hand side of Eq.~(\ref{eq:difbypart}) is then accounted for by defining 
\begin{equation}
\forcedensity^{\rm th} = \left( \kb T  \right) \gradposition\cdot \hydrospread_h.\label{eqn:thermforce}
\end{equation}
This thermal forcing term is discretized using the `random finite difference' (RFD) method described in Refs.~\citenum{delong2014brownian} and \citenum{ladiges2020discrete}. The second term on the right hand side of Eq.~(\ref{eq:difbypart}) is accounted for by the time stepping scheme described below in Sec.~\ref{subsec:TemporalAlgorithm}, which is also discussed in detail in Refs.~\citenum{delong2014brownian} and \citenum{ladiges2020discrete}.

The spreading and interpolation operators need to be modified when ions are close to a physical boundary, in particular when their support overlaps the boundary.
In a physically realistic simulation setup we would expect that a particle does not partially cross over a confining boundary; typically it would be repelled by some short range force such as those discussed in section \ref{sec:sr}. In particular, if the diameter of the support of the kernel used in the hydrodynamic spreading ($\hydrospread_h$) and interpolation ($\hydrointerp_h$) operators is similar to the van der Waals radius, we would expect that the kernels do not cross the boundary. However there are several situations where this is not the case. First, larger kernels such as the six point Peskin kernel \cite{New6ptKernel} may have support that exceeds the van der Waals radius so the edge of a the kernel can cross the boundary. Second, when a DISCOS simulation is being run with some degree of dry diffusion (see Section \ref{sec:drymob}), the support of the kernels grows relative to the true `size' of the particles, increasing the amount of overlap that can occur at a boundary. In each of these cases we must consider how to handle force spreading and velocity interpolation outside of the domain in a way that recovers physically realistic behavior.

\begin{figure}[h!]
\centering
    \includegraphics[width=0.7\textwidth]{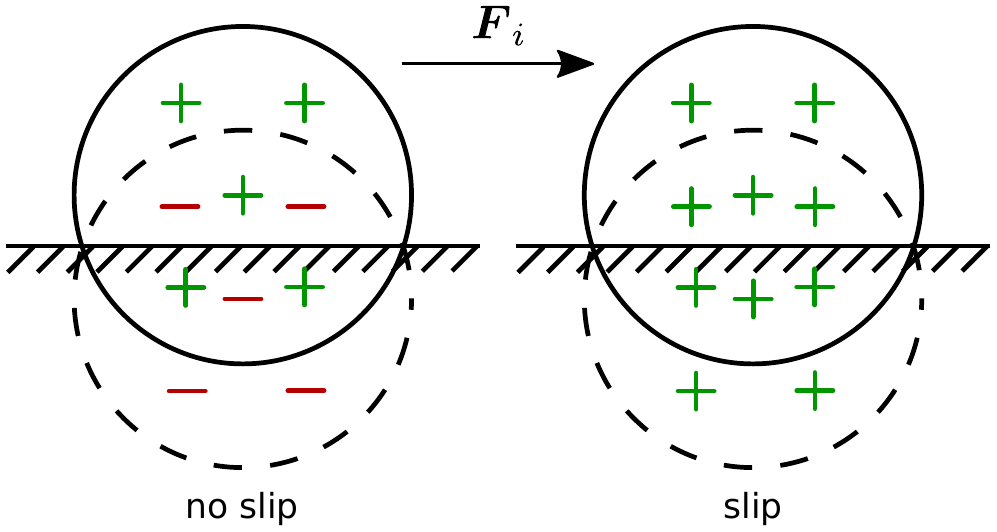}
 \caption{Illustration of the use of image particles when applying the spreading operation near a boundary. On the left, an image particle spreading the opposite tangential force will result in the particles tangential mobility going to zero as it approaches the boundary, the desired result for a no slip boundary. On the right, an image particle spreading the same tangential force will result in the unchanged tangential mobility expected from a full slip boundary condition. A separate spreading operation is performed for force normal to the boundary, for both types of boundary condition (Eq.~(\ref{fluidbc})) spreading the opposite force normal to the wall will result in zero normal mobility at the wall. As discussed later in Sec.~\ref{sec:electro}, this approach is also used when spreading charge for the electrostatic solution based on the use of either fixed potential or dielectric walls.
}
\label{fig:folded}
\end{figure}

We address this situation using a method of images. First, consider the spreading operator.
As the kernel representing the `real' particle leaves the domain, the kernel from the corresponding `image' particle will enter the domain as illustrated in Fig.~\ref{fig:folded}. 
For both boundary types discussed above, we wish the normal mobility of a particle to approach to zero as it approaches the wall. Therefore the image particle must spread an equal and opposite value for the normal force component. The same approach is used to recover zero tangential mobility for a no slip boundary. For the slip boundary we want the tangential mobility to remain unchanged. In this case the image particle must spread the same tangential force as the real particle for that component. 
These two conditions are illustrated on the left and right of Fig.~\ref{fig:folded}, respectively. We note that this approach is equivalent to that given in Refs.~\citenum{kallemov2016immersed}, where it is described without the use of image particles. The same image construction is also used in the Force Coupling Method for no-slip boundaries \cite{ForceCoupling_Channel} and also for slip boundaries \cite{FluctuatingFCM_DC}.

Analogously, to interpolate the fluid velocity on the grid to particle locations that are near the wall, ghost cells are included outside the domain filled such that the boundary condition is satisfied. The interpolation operation for kernels straddling the boundary may then be performed exactly as if the kernel were entirely within the domain. This perserves the adjoint condition Eq. (\ref{eq:adjoint}) in the presence of boundaries as discussed in Ref.~\citenum{delong2014brownian}.



\subsection{\label{subsec:Electrostatics}Particle forces}
We consider two types of forces acting on the particles:  long-range electrostatic forces and short-range forces. The net force on an ion is then
\begin{align}
\force_i = \force^{\mathrm{E}}_{i} + \sum_{j\in\Omega_i^{\mathrm{R}}}\force^{\mathrm{R}}_{ij} + \sum_{k\in\Omega_i^{\mathrm{W}}}\force^{\mathrm{W}}_{ik}+ \force^{\mathrm{ext}}_{i}. 
\label{eq:totalforce}
\end{align}
Here, $\force^\mathrm{E}_{i}$ is the electrostatic force and  $\force^\mathrm{R}_{ij}$ is the short-range force between particles $i$ and $j$, where
$\Omega_i^{\mathrm{R}}$ indicates all particles within a given range of the  $i^\mathrm{th}$ particle (see Eq.~\eqref{eq:potential} below). Additionally, $\force^{\mathrm{W}}_{ik}$ is the short range force between particle $i$ and wall $k$, and $\Omega_i^{\mathrm{W}}$ indicates all walls within a given range of particle $i$. The term $\force^\mathrm{ext}_{i}$ indicates forces due to an applied field, e.g., gravity or an external electric field.

\subsubsection{Electrostatic forces}\label{sec:electro}

For the computation of the electrostatic force, we will also use a particle and mesh approach.
However, unlike the hydrodynamics, in which the ions are represented as having non-zero size, for the electrostatic force computation we view the ions as point charges.
The charge density is
\begin{equation}
\varrho(\spatial) = \sum_{i=1}^N \delta(\position_i-\spatial) \charge_i,
\label{eq:delta_funcs}
\end{equation}
where $\delta$ is a Dirac delta function and $\chargevec=\{\charge_1,\cdots,\charge_i,\cdots,\charge_N\}$ are the ion charges.
Thus, the electrostatic force is found by solving Poisson's equation for the electrical potential $\phi$,
\begin{equation}
-\epsilon \gradspatial^2 \phi = \chargedensity,\label{eq:poisson}
\end{equation}
where  $\epsilon$ is the electrical permittivity of the solvent.

\textcolor{black}{For long range electrostatic interactions with domain boundaries, when solving Poisson's equation (Eq.~(\ref{eq:poisson})) we can specify a
fixed potential Dirichlet condition,  
\begin{align}
\phi(\spatial_b) = \phi_b\label{esbc1}
\end{align}
where $\phi_b$ is the potential on the wall.  A dialectric boundary with surface charge density $\varsigma$ is represented by the Neumann condition
\begin{align}
\left(\left.\epsilon \gradspatial \phi(\spatial)\right|_{\spatial\rightarrow\spatial_b^+}-\left.\epsilon_b \gradspatial \phi(\spatial)\right|_{\spatial\rightarrow\spatial_b^-}\right) = \varsigma \label{esbc2}
\end{align}
where $\epsilon_b$ is the permittivity of the wall, and $\spatial\rightarrow \spatial_b^+$ and $\spatial\rightarrow \spatial_b^-$ indicate the limit approaching the wall from the inward and outward normal directions, respectively. However in this paper we take the wall to be weakly polarizable compared to the solvent inside the domain, i.e., $\epsilon \gg \epsilon_b$. This yields the approximation
\begin{align}
\left. \gradspatial  \phi (\spatial) \cdot \wallnormal \right|_{\spatial=\spatial_b}=\varsigma/\epsilon,\label{esbc3}
\end{align}
This is typically a good approximation where the solvent is water. For an approach not constrained by this condition, see Ref.~\citenum{maxian2021fast}. Recent MD simulations \cite{cox2022dielectric} have also shown that in the case of water the simple model used here (treating the dielectric constant as a property constant across the thickness of the channel) is remarkably accurate in predicting the energy stored in a
parallel-plate capacitor down to nanometer-scale thickness, as long as
the dielectric boundary is properly placed at the location where the
density of hydrogen approximately vanishes. Note that both $\phi_b$ and $\varsigma$ can be spatially-varying, and different regions on the same boundary can be specified as either Dirichlet or Neumann. In a system with no Dirichlet boundaries, i.e. all Neumann or mixed Neumann and periodic, the total system must be electro-neutral. That is, the total surface charge and total charge within the domain must sum to zero.}

The electric field is obtained from the solution of Eq.~(\ref{eq:poisson}) with
$\efield =- \bs \nabla_r \phi$. An electrostatic kernel $\electrokernel$ is used to define electrostatic discrete spreading
and interpolation operators $\electrospread_h$ and $\electrointerp_h$ analogous to Eq.~(\ref{kernel2}) and Eq.~(\ref{kernel1}). In this case the charges $\chargevec$ are spread to the grid to give the charge density $\varrho$, and the resulting electric field is interpolated back to the particle locations using
\begin{align}
\efield^{\rm P} = \electrointerp_h \left(-\bs\nabla_h (-\epsilon \nabla_h^2 )^{-1} \electrospread_h \chargevec\right), \label{eq:efieldop}
\end{align}
where $\efield^{\rm P} = \left\{\efield^{\rm P}_1,\cdots,\efield^{\rm P}_i\cdots\efield^{\rm P}_N\right\}$ is the electric field at each particle location. For particle $i$, the associated force is then $\force^\mathrm{P}_i = \charge_i \efield_i^{\rm P}$. We note that the operators $\electrospread_h$ and $\electrointerp_h$ need not use the same kernels as their hydrodynamic counterparts, and Eq.~(\ref{eq:poisson}) need not be solved on the same grid.

As was the case for the hydrodynamic interactions, when ions are near physical boundaries, we need to modify the
computation of the electric field.
\textcolor{black}{Similar to the hydrodynamic solution, the image construction is implemented to produce the desired behavior as a particle approaches a wall. For a Dirichlet (fixed normal derivative) boundary, spreading an image of opposite charge leaves the potential constant on the boundary. For a Neumann (fixed potential gradient), spreading an image of like charge leaves the derivative of the potential normal to the boundary unchanged.} As with the velocity in the hydrodynamic case, sufficient ghost cells for the resulting electric field ($\efield^\text{P}$) must be filled for the support of the kernel used in the interpolation operator $\electrointerp_h$; these are filled such that the boundary condition on $\phi$ is satisfied. 

As noted above, we wish to treat the ions as point particles. As a consequence the electric field, and therefore the force, between particles with separation comparable to the mesh spacing will not be accurately resolved. To address this issue, we compute the overall electrostatic force using a particle-particle, particle-mesh (P3M) method \cite{Hockney:1988:CSU:62815, frenkel2001understanding}. The above provides an accurate representation of the electrostatic force at ranges that are large with respect to the mesh spacing. To resolve shorter ranges we use the direct Coulomb force between charged particles, 
\begin{align}
    \force_{ij}^\mathrm{C}= \frac{1}{4\pi \epsilon} \frac{\charge_i \charge_j}{\absposition_{ij}^2}\hat{\position}_{ij},\label{eq:coulomb}
\end{align}
where $\position_{ij}=\position_i-\position_j$, $\absposition_{ij} = |\position_{ij}|$, and  $\hat{\position}_{ij} = \position_{ij}/\absposition_{ij}$ is the unit vector.
We introduce a local correction for nearby particles so that
\begin{align}
    \force_{i}^\mathrm{E}= \force^\mathrm{P}_i +  \sum_{j\in\Omega_i^\mathrm{E}} \force_{ij}^\mathrm{LC},\label{eq:fullElectro}
\end{align}
where $j\in\Omega_i^\mathrm{E}$ represents all particles within short range cuttoff distance $\psi$ of particle $i$, and 
\begin{align}
   \force_{ij}^\mathrm{LC} = \force_{ij}^\mathrm{C} - \force_{ij}^\mathrm{P}\label{eq:LC}. 
\end{align}
\textcolor{black}{Here $\force_{ij}^\mathrm{P}$ is the force computed on the grid for two nearby particles in an unbounded domain, which
can be efficiently pre-tabulated as a function of particle separations as described in Ref.~\citenum{ladiges2020discrete}.}

For particles near the boundary, we perform the near-field correction including the image particles described above when a particle is within distance $\cutoff/2$ from the boundary, i.e., within the short range cut off distance $\cutoff$ of its own image charge. Note that the correction must also be applied for image charges from other particles that fall within range $\cutoff$. For simplicity, in the results given in the following section we have set $\cutoff$ to be equal to the kernel radius. Note that this is not a requirement, the optimal choice of $\cutoff$ is discussed in Appendix 2 of Ref.~\citenum{ladiges2020discrete}.

\subsubsection{Short range forces}\label{sec:sr}

In addition to electrostatic forces, we incorporate a short range force to account for quantum effects, such as Fermi exclusion, that prevent ions of opposite charge from getting too close.
The short range interactions are specified using a potential function $U^{\rm sr}(\paramposition;\sigma,\xi)$, where $\paramposition$ is the radial distance from a particle, $\sigma$ is the van der Waals diameter, and $\xi$ is the magnitude of the potential. In Ref.~\citenum{ladiges2020discrete} we employed the repulsive only Weeks-Chandler-Andersen (WCA) potential \cite{Weeks1971,Hockney:1988:CSU:62815};  in Section \ref{sec:Results} we also make use of the complete Lennard-Jones potential \cite{jones1924determination1,jones1924determination2,frenkel2001understanding}
\begin{align}
U^{\rm sr}(\paramposition;\sigma,\xi) =
\begin{cases}
\widehat{U}^{\rm sr}(\paramposition;\sigma,\xi) - \widehat{U}^{\rm sr}(\paramposition_c;\sigma,\xi), \vspace{1mm}&0 < \paramposition < \paramposition_c \\
0, & \paramposition_c \ge \paramposition
\end{cases},\label{eq:potential}
\end{align}
where
\begin{align}
\widehat{U}^{\rm sr}(\paramposition;\sigma,\xi) = 4\xi \left(\left(\frac{\displaystyle\sigma}{\displaystyle\paramposition}\right)^{12} - \left(\frac{\displaystyle\sigma}{\displaystyle\paramposition}\right)^6\right).
\end{align}
The WCA potential is recovered by setting the cutoff distance $\paramposition_c = 2^{1/6}\sigma$. When using the complete Lennard-Jones potential we have set $\paramposition_c = 2.5\sigma$. Forces are then calculated pairwise using
\begin{equation}
\force^\mathrm{R}_{ij} = -\hat{\position}_{ij}\,\frac{d}{d \paramposition} U^{\rm sr}_{ij}(\absposition_{ij}; \sigma_{ij},\xi_{ij}),\label{eq:closerange}
\end{equation}
where $\sigma_{ij}$ and $\xi_{ij}$ are specified for each pair of particles $i$ and $j$.

Similarly, we include short range forces for particle interactions with walls. The treatment of short range forces in this case is straightforward; a potential function is defined for particle interactions with a wall and the resulting force applied to any particles within range of the wall. There are a number of possible potential functions, the complex atomic structure of a wall may, for example, be treated as a `soft' potential. The interaction of this structure with a hard sphere particle is approximated by the 9-3 potential \cite{abraham1977structure}
\begin{align}\label{eq:LJ93}
\widehat{U}^{\rm sr}(\paramposition;\sigma,\xi) = \frac{3^{3/2}}{2}\xi \left(\left(\frac{\displaystyle\sigma}{\displaystyle\paramposition}\right)^{9} - \left(\frac{\displaystyle\sigma}{\displaystyle\paramposition}\right)^3\right).
\end{align}
from which $\force^{\mathrm{W}}_{ik}$ is computed. In the following sections where we have employed the 9-3 potential, a cutoff of $\paramposition_c = 3.5\sigma$ is used.

\subsection{Dry diffusion}\label{sec:drymob}


As discused in Section~\ref{sec:BD}, compact kernel functions are used to define particle hydrodynamic interactions. Each of these kernel functions (see Appendix \ref{appdx2}) depend on the cell size, $\Delta r$, which makes their effective hydrodynamic radius $\radius_w$ vary with grid resolution.
From the Stokes-Einstein relation, this leads to a diffusion coefficient
\begin{equation}
D^{\rm wet} = \frac{\kb T}{b \eta \radius_w}.
\end{equation}
that depends on both $\Delta r$ and the form of the kernel function.
Here, $b$ is a constant given by the boundary condition on the particle; we take $b = 6\pi$, \textcolor{black}{corresponding to a no-slip boundary condition on the surface of a spherical particle.}

If the actual hydrodynamic radius of the ion, $\radius_t$, is smaller than $\radius_w$ then the diffusion coefficient, $D^{\rm wet}$, will be too small. 
As a correction, we add a `dry' diffusion term
\begin{equation}
D^{\rm dry} = \frac{\kb T}
{b \eta \radius_d},\label{einstein}
\end{equation}
such that the total diffusion is
\begin{equation}
D^{\rm tot} = D^{\rm wet} + D^{\rm dry} \label{diffSum}.
\end{equation}
The hydrodynamic radius corresponding to the dry diffusion can be computed from
$\radius_t = (\radius_w^{-1} + \radius_d^{-1})^{-1}$.
This approach makes it possible to use a coarser grid for the Stokes solver than
would be dictated by the hydrodynamic radius of the ions, compensating for the lower resolution with the dry diffusion. This yields improved computational performance by allowing a coarser grid, at the cost of neglecting short range hydrodynamic interactions. 
The effect of this trade-off is examined in detail in Ref.~\citenum{ladiges2020discrete}, and below in Sec.~\ref{sec:result_eo}.

When an ion gets close to a boundary, it mobility is reduced.
The spreading and interpolation procedures described above in Section~\ref{sec:BD} produce the desired reduction of mobility in the vicinity of a wall for the wet part of the particle diffusion. 
However, we also need to account for the reduction in dry mobility induced by the boundary.
To incorporate this effect we define a dry diffusion tensor, which is a function of particle position,
\begin{align}
\bs{D}^\text{dry}_i = D_i^\text{dry}\bs{\Gamma}_i(\position_i),\label{eqn:diff1}
\end{align}
where $D^\text{dry}_i$ is the value for a single particle in an infinite domain, and
\begin{align}
\bs{\Gamma}_i(\position_i) = \frac{\bs{\gamma}^{\rm tot}_i(\position_i)D_i^{\rm tot} -  \bs{\gamma}^{\rm wet}_i(\position_i)D_i^{\rm wet}} {D_i^{\rm dry}}.\label{eqn:diff2}
\end{align}
is a vector quantity to allow for anisotropic mobility.

The functions $\bs{\gamma}_i$ are vector quantities with elements between zero and one representing the modification to the free diffusion. For example, the total diffusion of freely diffusing particle is $D_i^{\rm tot}$, and when confined by boundaries it is given by $\bs{\gamma}^{\rm tot}_i D_i^{\rm tot}$. A particle sufficiently far from any boundaries would have $\bs{\gamma}^{\rm tot}_i = (1,1,1)$, and a particle where $\position_i$ is coincident with a no-slip boundary would have $\bs{\gamma}^{\rm tot}_i = (0,0,0)$. \textcolor{black}{For a given particle the functions $\bs{\gamma}_i$ depend on the hydrodynamic radius, which in turn depends on the kernel type and cell size, however they do not apply at distances very close to the boundary.} Analytic expressions exist for $\bs{\gamma}_i$ in the case of several simple geometries and kernels \cite{swan2007simulation,swan2010particle,aponte2016simulation}. \textcolor{black}{However here we take the general approach of numerically pre-computing by measuring the particle diffusion at multiple locations from which the functions can be interpolated or fit in subsequent simulations employing that geometry.}
The function $\bs{\gamma}^{\rm tot}_i$ is measured using a cell size corresponding to a 100\% wet simulation, $\bs{\gamma}^{\rm wet}_i$ is measured on a grid producing the wet percentage the simulation will be run at. Examples are given in Section~\ref{sec:result_drymob}.

Incorporating both the dry diffusion terms discussed here, and the wet components discussed in Section~\ref{sec:BD}, the total equation of motion for a particle is
\begin{equation}
\textcolor{black}{\frac{d \position_i}{d t} = \underbracket[0.4pt][2pt]{\bs{V}_i \vphantom{ \frac{D_i^{\rm dry}}{\kb T}\force_i }       }_{\text{wet}} 
+ \underbracket[0.4pt][2pt]{ \frac{\bs{D}_i^\text{dry}}{\kb T}\force_i +  \gradposition \cdot \bs{D}_i^\text{dry}  +  (2 \bs{D}_i^\text{dry})^{1/2}\ito\particlenoise_i^{\rm dry}}_{\text{dry}},}
\label{eq:ions3}
\end{equation}
\textcolor{black}{where $\ito\particlenoise_i^{\rm dry}$ is a standard white noise process and $ \gradposition \cdot \bs{D}_i^\text{dry}$ is the It\^{o} stochastic drift term corresponding to the dry diffusion. If the functional form of $\bs{D}_i^\text{dry}$ is known this term may be computed directly. Otherwise, it can be approximated using finite differences, or by differentiating an interpolant of $\bs{D}_i^\text{dry}$.}


\subsection{Temporal Algorithm} \label{subsec:TemporalAlgorithm}
\textcolor{black}{The time step is defined by the following four stages:}
\begin{enumerate}

\item The charge density $\varrho$ is spread to the grid using the spreading operation, $\electrospread_h \chargevec$.
Poisson's equation, Eq.~(\ref{eq:poisson}), is then solved using the geometric multigrid method to obtain the coarse electric field, $\efield = - \bs \nabla_r \phi$.

\item This electric field is mapped to the particle locations using the interpolation operation,
$\efield^\text{P}=\electrointerp_h \efield$,
and the corresponding force on the particles is computed using $\force^\mathrm{P}_i = \charge_i \efield_i^{\rm P}$. For particles at close range, this force is corrected using the P3M approach per Eqs.~(\ref{eq:coulomb}), (\ref{eq:fullElectro}), and (\ref{eq:LC}). Short range forces are computed using Eq.~(\ref{eq:closerange}), and the total force on the particle is calculated as per Eq.~(\ref{eq:totalforce}).
\item \textcolor{black}{The force on the particles, $\force$, is mapped to the grid 
using the spreading operation defined in Eq.~(\ref{kernel2}) to compute $\forcedensity^\mathrm{s}$.
The resulting linear system,
\begin{subequations}
\begin{align}
\graddiscrete p - \eta \graddiscrete^2 \fluidvel =&\, \forcedensity^\text{s} + \forcedensity^\mathrm{th}+\sqrt{\frac{2 \kb T \eta}{\cellvol \Delta t }}\, \graddiscrete\cdot \fulldiscretenoise,\label{stokes_temporal}\\
\graddiscrete \cdot  \bs v =&\, 0,
\end{align}\label{eq:disc_stokes_temp}%
\end{subequations} 
is solved using preconditioned GMRES \cite{nonaka2015low} to compute the velocity at time step $n$, $\fluidvel^n$. The thermal forcing term $\forcedensity^\mathrm{th}$ is included using the random finite difference method. Note that $\fulldiscretenoise$ is a collection of independent Gaussian random numbers of variance one that have been scaled by $1/\sqrt{\Delta t}$ to represent the time discretization of $\discretefieldnoise$. }
\item \textcolor{black}{The fluid velocity $\fluidvel^n$ is mapped to particle locations using the interpolation operator defined in Eq.~(\ref{kernel1}) to obtain the ``wet'' component of the particle velocities. The temporal discretization of the particle location update is then given by the midpoint update scheme
\begin{subequations}
\begin{eqnarray}
\position_i^{n+1/2,\star} &=& \position_i^n + \frac{\Delta t}{2}(\hydrointerp_h(\position_i^n)\fluidvel^n), \label{eq:step1}\\
\position_i^{n+1} &=& \position_i^n+\Delta t \left[\hydrointerp_h(\position_i^{n+1/2,\star})\fluidvel^n + \frac{D_i^{\rm dry}}{\kb T}\force_i^n
+ \gradposition \cdot \bs{D}_i^\text{dry}   \right.
+\left. \sqrt{\frac{2D_i^{\rm dry}}{\Delta t}} \bs{W}_i^{\rm dry,n}\right].
\label{eq:step2}
\end{eqnarray}
\label{fixman}%
\end{subequations}
where $\discreteparticlenoise_i^{\rm dry, n}$ are independent Gaussian random numbers of variance one. As discussed in detail in Ref.~\citenum{delong2014brownian}, the midpoint update incorporates the second part of the stochastic drift term in Eq.~(\ref{eq:difbypart}) that is not accounted for by the random finite difference force density $\forcedensity^\mathrm{th}$. See Section~IID and Appendix A in Ref.~\citenum{delong2014brownian}.}
\end{enumerate}

\section{Numerical Tests}\label{sec:Results}

\begin{figure}[h!]
  \centering
    \includegraphics[width=0.45\textwidth]{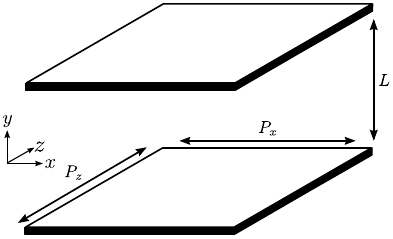}
 \caption{Problem geometry. Each of the test cases given below utilizes a channel of height $L$, which is periodic with width and depth $P_x$ and $P_z$, respectively; each of these parameters is specified in the relevant section.}
\label{fig:geometry}
\end{figure}

In this section we perform several numerical validation tests of the DISCOS algorithm for an electrolyte confined in a channel.
In Sec.~\ref{sec:result_drymob} we examine the approach described in Sec.~\ref{sec:drymob} for incorporating changes to the dry mobility due to the channel walls.
In Sec.~\ref{sec:result_esbc} we test the method for implementing electrostatic boundary conditions given in
Sec.~\ref{sec:electro}.
In Sec.~\ref{sec:result_eq} we compare the ion distribution at equilibrium to existing numerical results.
In Sec.~\ref{sec:result_stochdrift} we use this example to test the application of both the wet and dry stochastic drift terms discussed in Sec.~\ref{sec:drymob}. Each of these test cases involves a channel confined in the $y$ direction, with periodic boundary conditions in the $x$ and $z$ directions, as illustrated in Fig.~\ref{fig:geometry}.
In all cases a no-slip boundary condition is used for the hydrodynamic solution. The electrostatic configuration, and additional parameters specific to each problem, are described in the corresponding section.

\subsection{Particle mobility}\label{sec:result_drymob}
As discussed  in Sec.~\ref{sec:drymob}, in confined systems we need to determine the correction to the dry mobility of the ions to account for the presence of no-slip walls that impart drag on the fluid. This requires finding the mobility reduction functions $\bs{\gamma}_i$. As per  Eq.~(\ref{eqn:diff2}), we must measure this function for both the total and wet hydrodynamic radii, $a_t$ and $a_w$, giving $\bs{\gamma}^{\rm total}_i$ and $\bs{\gamma}^{\rm wet}_i$, respectively. Note that these values may differ for each species of ion in the electrolyte, in which case $\bs{\gamma}_i$ must be measured separately for each species.
As mentioned above, analytic formulas exist for some simple geometries. For an infinite channel of height $L$, where $L \gg \radius$ we have \cite{swan2010particle} 
\begin{align}
\gamma_i(\tilde{y}) \approx \left\{ 1+\sum_{n=0}^\infty (-1)^n \left[ \frac{1}{\gamma^{*}\left( n L + \tilde{y},\radius\right)}-1 \right]+\sum_{n=0}^\infty (-1)^n \left[ \frac{1}{\gamma^{*}\left( (n+1) L - \tilde{y},\radius\right)}-1 \right]\right\},\label{mob_analytic}
\end{align}
where $\gamma_i$ is defined for parallel, $\gamma_{\parallel i}$, and perpendicular, $\gamma_{\bot i}$, to the channel wall, with $\tilde{y}$ giving the distance to the nearest wall. In both cases $\gamma_i^*$ is given by the corresponding formula for a single infinite plane,
\begin{align}
\gamma^*_{\parallel i}(\tilde{y}) \approx 1 - \frac{9 a}{16 \tilde{y}} + \frac{2 a}{16 \tilde{y}^3}- \frac{ a}{16 \tilde{y}^5}, \qquad \gamma^*_{\bot i}(\tilde{y}) \approx 1 - \frac{9 a}{8 \tilde{y}} + \frac{4 a}{8 \tilde{y}^3}- \frac{ a}{8 \tilde{y}^5}.\label{singlewall}
\end{align}
Equation~(\ref{mob_analytic}) is sufficiently accurate for our purposes so long as the particle is at least approximately\footnote{This limitation arises from the order of expansion used to derive Eq.~(\ref{singlewall}).} 0.7 radii from the wall. For a simulation that is completely wet, i.e. $a_w=a_t$, this would normally be the case due to short range repulsion (see Eq. (\ref{eq:LJ93})). However for a simulation featuring dry diffusion, e.g. a 50\% wet $a_w=a_d=2 a_t$, a particle can approach a wall to within a distance less than $a_w$. In this case Eq.~(\ref{mob_analytic}) can not be used to find $\bs{\gamma}_i^{\rm wet}$, which we note again is required to compute Eq.~(\ref{eqn:diff2}).

\begin{figure}[h!]
  \centering
    \includegraphics[width=0.95\textwidth]{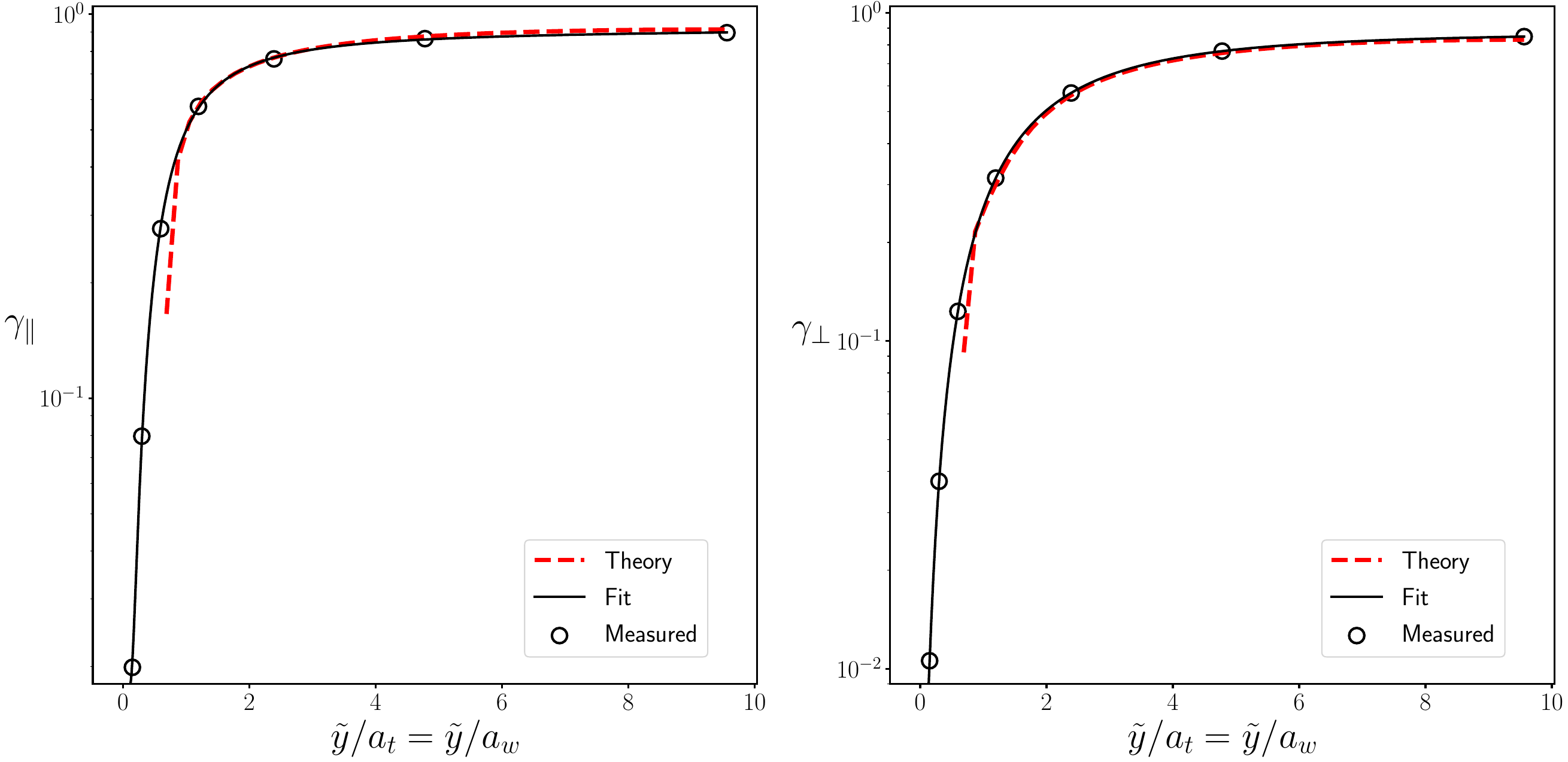}
 \caption{Correction to mobility (or diffusion) as a function of distance to the nearest channel wall. The circles indicate the values measured from a fully wet simulation, i.e., $D^\text{dry}=0$ or $a_d\rightarrow \infty$; note that the point $(0,0)$ is also included in the set. The solid black line represents a least squares fit, Eq.~(\ref{eq:fitfunction}), to this data; this is the function actually used in subsequent simulations. The dashed red line is given by Eq.~(\ref{mob_analytic}); note its divergence at distances less than 0.7 radii. 
 }
\label{fig:mobfit1}
\end{figure}

\FloatBarrier

Due to this issue, and as in general more complex geometries will not have analytic approximations, we directly measure the functions $\bs{\gamma}_i$ for this channel; where possible we use Eq.~(\ref{mob_analytic}) for comparison. \textcolor{black}{This is done by placing a particle at a given location, setting $\kb T=0$, applying a force, and noting the resulting particle velocity.} Figure~\ref{fig:mobfit1} shows the values measured for a channel with no-slip boundary conditions in the $y$ direction, for $L=3\text{nm}$, and periodic boundaries in $x$ and $z$, where $P_x=P_z=12\text{nm}$. This is also the channel size used Sec.~\ref{sec:result_eq}. Note that increasing $P_x$ and $P_z$ does not significantly change any of the results given in this section, the interaction with the walls dominates any periodic effect on the particle mobility. The desired particle size is $a_t=0.157\text{nm}$, which is achieved with the 4-point Peskin kernel when $\Delta r = 0.25\text{nm}$ (24 cells are used to represent the channel height). For the region where Eq.~(\ref{mob_analytic}) is valid we see good agreement with theory. To represent the measured data in subsequent simulations we use a fit function of the form
\begin{align}
\hat{\gamma}_i(\tilde{y}) = c_1 + \frac{c_2}{(c_3 + \tilde{y})} + \frac{c_4}{(c_5 + \tilde{y})^3} + \frac{c_6}{(c_7 + \tilde{y})^5},\label{eq:fitfunction}
\end{align}
where $c_1$-$c_7$ are the fit parameters. This is also shown in Fig.~\ref{fig:mobfit1}. 

\begin{figure}[h!]
  \centering
    \includegraphics[width=0.95\textwidth]{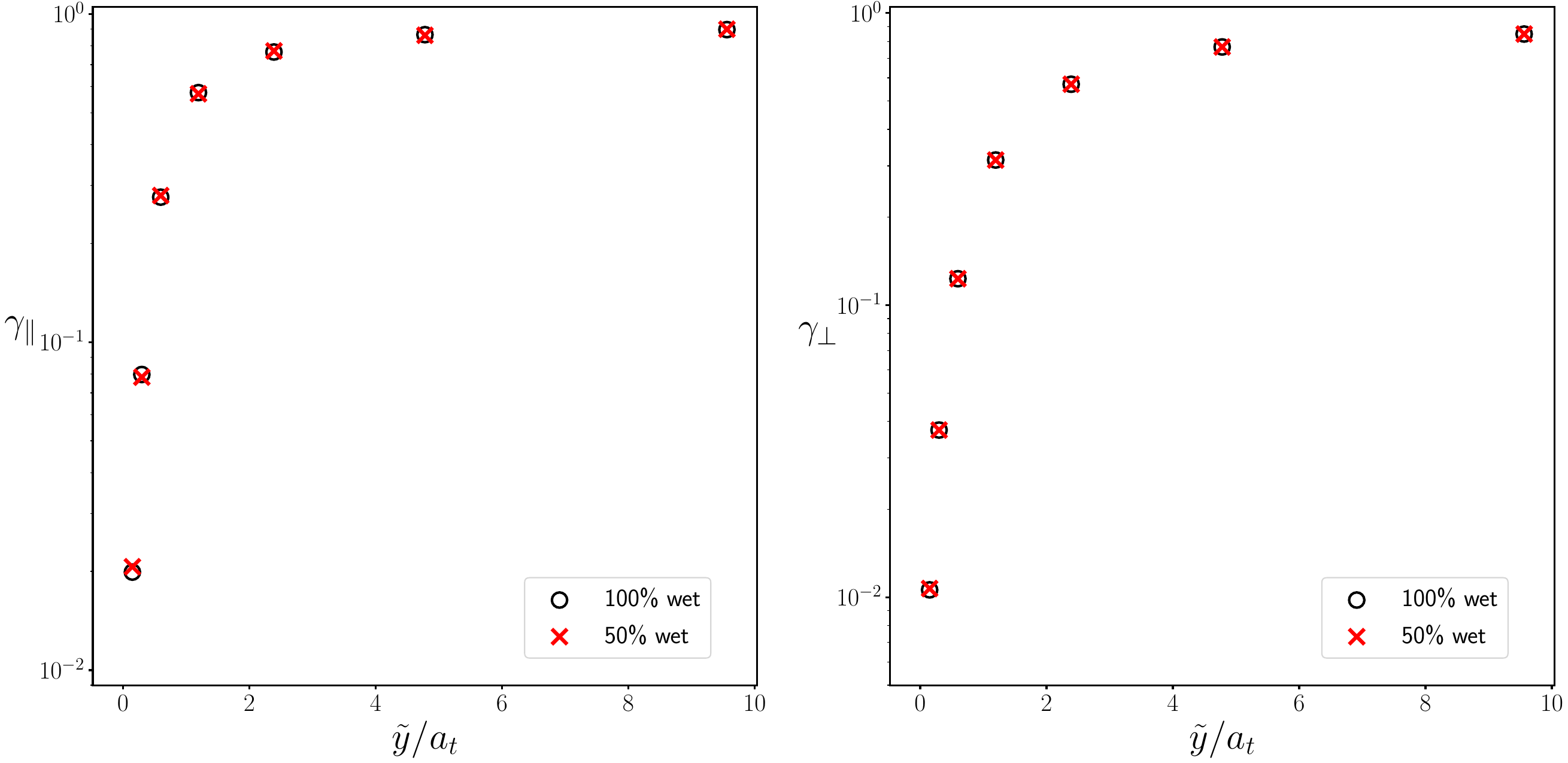}
 \caption{Comparison of particle mobility in 100\% and 50\% wet simulations using the method described in sections \ref{sec:drymob} and \ref{sec:result_drymob}.}
\label{fig:mobfit3}
\end{figure}

To perform simulations of a given wet percentage we are required to determine $\bs{\gamma}^{\rm total}_i$ and $\bs{\gamma}^{\rm wet}_i$ and use those functions in Eq.~(\ref{eqn:diff2}) to find the total particle diffusion. Taking the above parameters to represent $\bs{\gamma}^{\rm total}_i$, we may obtain $\bs{\gamma}_i^{\rm wet}$ for a 50\% wet simulation by making the same measurements with the cell size doubled, or alternatively, on a channel of height $L/2$ while keeping the cell size the same. In Fig.~\ref{fig:mobfit3} the resulting $\bs{\Gamma}_i$ is compared between 50\% and 100\% wet ($\bs{\gamma}_i^{\rm wet} = \bs{\gamma}_i^{\rm total}$) simulations, and good agreement is observed; slight discrepancies are the result of translational variance in the Peskin kernels.

\begin{figure}[h!]
  \centering
    \includegraphics[width=0.95\textwidth]{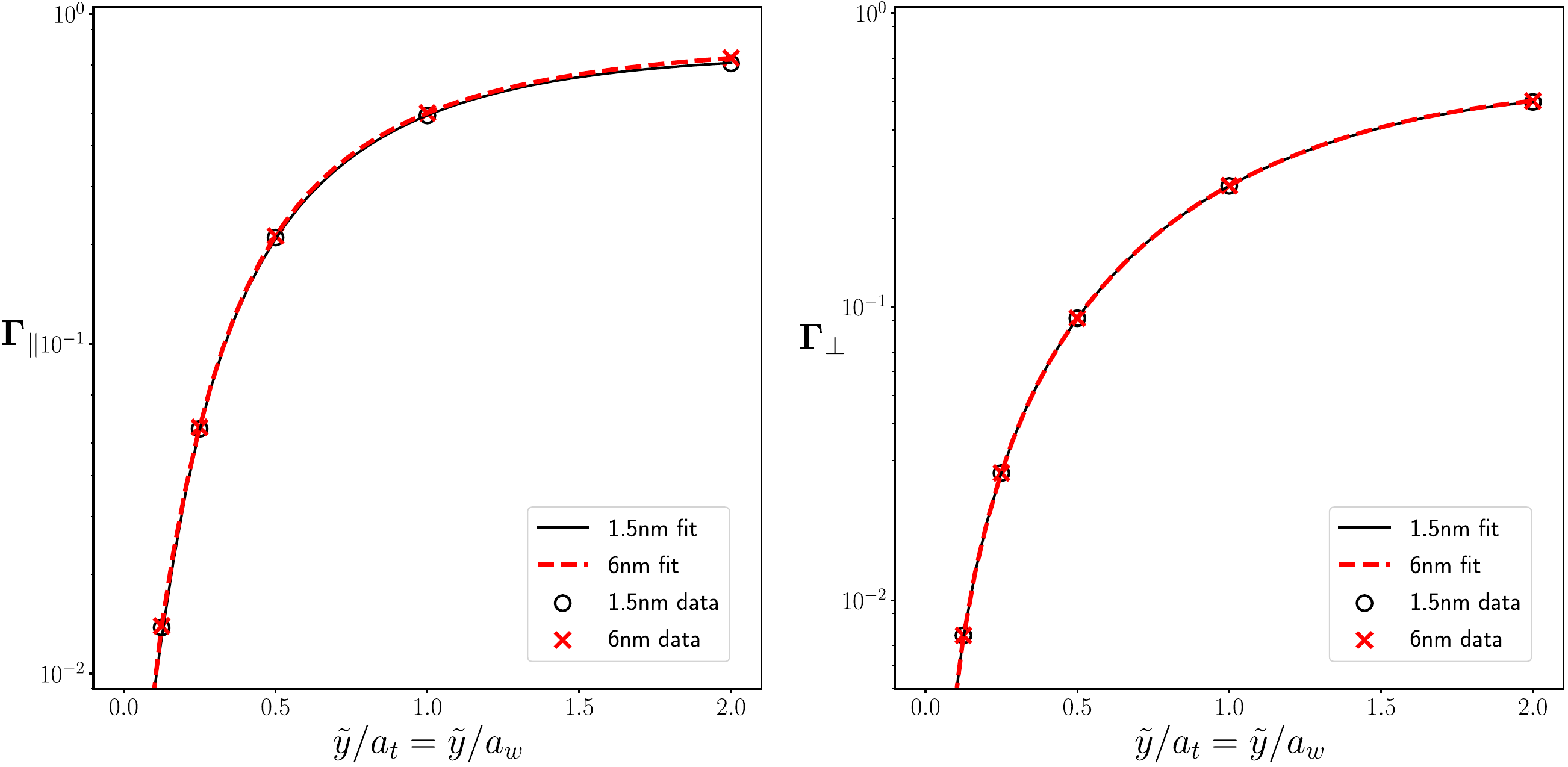}
 \caption{Comparison of near wall mobility for different channels, illustrating that for sufficiently large channels the near wall mobility may be described by a common function.}
\label{fig:mobfit2}
\end{figure}

An alternative approach (not used here) for simulating an arbitrary channel is to employ Eq.~(\ref{mob_analytic}) for the region $\tilde{y}>a_w$, and Eq.~(\ref{eq:fitfunction}) for $\tilde{y}<a_w$. As shown in Fig.~\ref{fig:mobfit2}, for a sufficiently large channel, hydrodynamics in the region $\tilde{y}<a$ are dominated by the nearest wall, removing the length scale $L$ from consideration. In this region a single measurement of Eq.~(\ref{eq:fitfunction}) can be applied to any sufficiently large channel (see Fig.~\ref{fig:mobfit2}).

\subsection{Electrostatic boundary conditions}\label{sec:result_esbc}

Here we test the P3M method described in sections \ref{sec:electro} 
for implementing electrostatic boundary conditions. We employ a channel with $L=6\text{nm}$ and $P_x=P_z=48\text{nm}$, and a cell size of $\Delta r = 0.09375\text{nm}$, i.e., 64 cells in the $y$ direction. Two ions with $q=1.6\times 10^{-19} \text{C}$ are placed at a distance $2\Delta r$ from each other in the $x-z$ plane. The absolute force on each ion is then measured at varying distances from the wall. For each distance the ions are placed at a fixed $y$, and one hundred different measurements are averaged at random $x$ and $z$ locations (while maintaining $2\Delta r$ separation) to account for the slight translational variance of the kernels. We note that a small error will still be present due to keeping $y$ fixed; see Ref.~\citenum{ladiges2020discrete} for further discussion. The four point kernel was used, so the ions are in the P3M correction range ($\psi$) of each other, and when sufficiently close to the wall each ion will be in the correction range of both it's own image charge and that of the other ion. Two cases were tested, homogeneous Dirichlet, and inhomogeneous Neumann, i.e., 
\begin{align*}
\phi(y=0) = \phi(y=L) = 0, \quad \text{and} \quad  \left. \frac{\partial \phi}{\partial y} \right |_{y=0} = \left.-\frac{\partial \phi}{\partial y} \right |_{y=L} =\varsigma/\epsilon,
\end{align*}
where $\varsigma= -6.94\times10^{-9}\mathrm{C/cm}^2$
to give an electro-neutral system. For both case we have used $\epsilon_r = 66.3$,\footnote{This value of $\epsilon_r$ is used in electro-osmosis simulations in Sec.~\ref{sec:result_eo}, for convenience we have used it here also.}
where $\epsilon = \epsilon_r \epsilon_0$, and $\epsilon_0$ is the vacuum permittivity.

\begin{figure}[h!]
  \centering
    \includegraphics[width=0.95\textwidth]{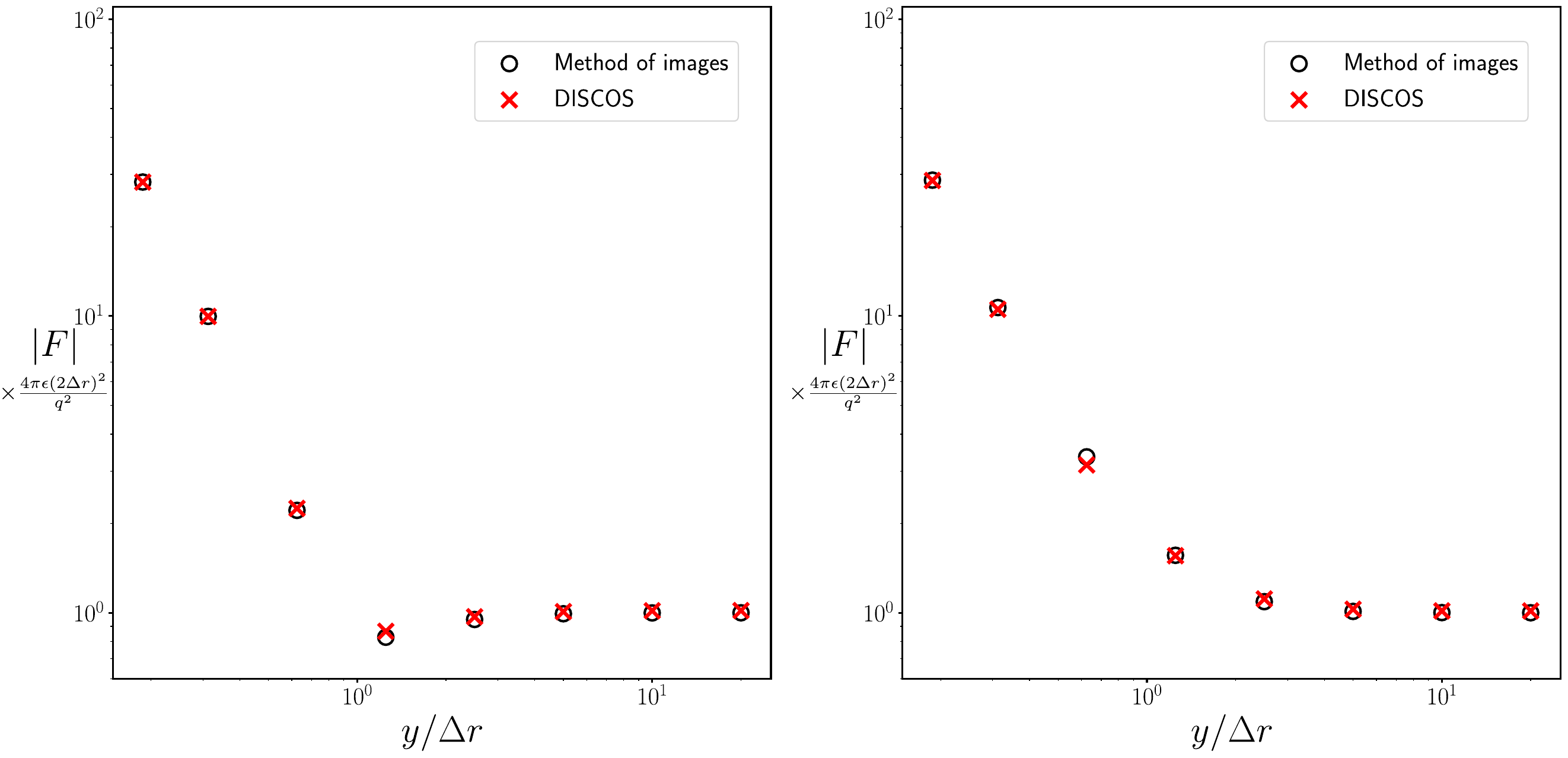}
 \caption{Absolute force on an ion in a two particle system at varying distances from the channel wall. Left: Homogeneous Dirichlet condition. Right: Inhomogeneous Neumann condition. Black circles indicate the forces calculated using the method of images, red crosses are those found using the P3M method described in sections \ref{sec:electro}
. }
\label{fig:electrores}
\end{figure}

In Fig.~\ref{fig:electrores} the absolute force on an ion is compared to that calculated using the method of images; in each case 500 images were used ensuring convergence to a higher accuracy than the P3M solution. In all cases good agreement is observed.

\subsection{Equilibrium in a confined channel}\label{sec:result_eq}

Here we test the DISCOS algorithm in reproducing the correct equilibrium distributions of ions in a confined channel by comparing with results given in Ref.~\citenum{jing2015ionic}. In that paper a channel height of $L=3\text{nm}$ was examined, containing anions and cations of equal total charge where the cation concentration was 0.5M; this corresponds to a Debye length of $\Dlength=3.9\text{nm}$ (see Eq.~(\ref{lengths})). In Ref.~\citenum{jing2015ionic} the walls are taken to have the same permittivity as the water inside the channel, and external electric field corresponding to a potential difference of 0.1V was applied perpendicular to the channel, causing an asymmetrical distribution of cations and anions.
\textcolor{black}{This is equivalent to setting a boundary condition in the far field when solving Eq.~(\ref{eq:poisson}), however for this problem it is feasible to take the more straightforward approach of performing a direct pairwise calculation of the electrostatic interactions \textcolor{black}{(including periodic images in the $x$ and $z$ directions)} using the Coulomb equation, Eq.~(\ref{eq:coulomb}). Following Ref.~\citenum{jing2015ionic}, we have used relative permittivity $\epsilon_r = 80$. The simulations in the referenced paper employed an implicit solvent, so at equilibrium we expect essentially identical results to those from the DISCOS simulations.}

To perform 100\% wet simulations 24 grid cells were used in the $y$ direction, with 96 cells in $x$ and $z$ ($P_x=P_y=12\text{nm}$). 
Setting $\eta=0.01\mathrm{g/(cm \, s)}$, this corresponds to a diffusion coefficient of $1.3915\times10^{-5}\text{cm}^2/\text{s}$ for the 4-point Peskin kernel. The 50\% wet simulations were performed by halving the number of cells in each dimension; for these simulations we have used the functions $\bs{\gamma}_i(\radius_t)$ and $\bs{\gamma}_i(\radius_w)$ measured in the previous section. For systems at equilibrium we expect the result to be independent of particle mobility, and therefore insensitive to these numerical parameters. However, as we see in Section~\ref{sec:result_stochdrift}, they provide a useful avenue to test the accuracy of the stochastic drift implementation. 

The diffusive timescale, $\radius^2_t /D^\text{tot}$, for the ions is on the order of 10ps. The time step used in the DISCOS simulations is constrained by this timescale, and also by the stiffness of the short range and electrostatic interactions. The importance of these interactions is somewhat
dependent on the local concentrations of ions which are problem dependent. For simplicity, rather than determining the optimal value in each case, throughout this paper we have employed a conservative time step of $\Delta t=0.1\text{ps}$; in principle larger time steps could be used in many of the cases examined, although these would still need to be substantially below the diffusive timescale.

Two methods are used in Ref.~\citenum{jing2015ionic} to determine the ion distribution: solving the Ornstein-Zernike (OZ) equation with the anisotropic hypernetted chain closure (AHNC) \cite{kjellander1984correlation, kjellander1985inhomogeneous}, and an implicit solvent molecular dynamics (MD) approach. The principal difference between the two is that the OZ solution contains genuine hard sphere inter-particle potentials, while the MD simulations employ purely repulsive Lennard-Jones (WCA) interactions (see Sec.~\ref{sec:sr}). In each case a van der Waals diameter of $\sigma=0.714\text{nm}$ is used. It is unclear what magnitude is used for the LJ interactions in the MD simulations, so here we have used a relatively large $\xi=4\kb T \approx 4\times10^{-21}\text{J}$; this is sufficiently close to genuine hard sphere that we can compare our results to the OZ/AHNC solution. Note that a purely repulsive 12-6 Lennard-Jones potential is still used for wall interactions, in this case we take $\sigma=0.357\text{nm}$.

\begin{figure}[h!]
  \centering
    \includegraphics[width=0.95\textwidth]{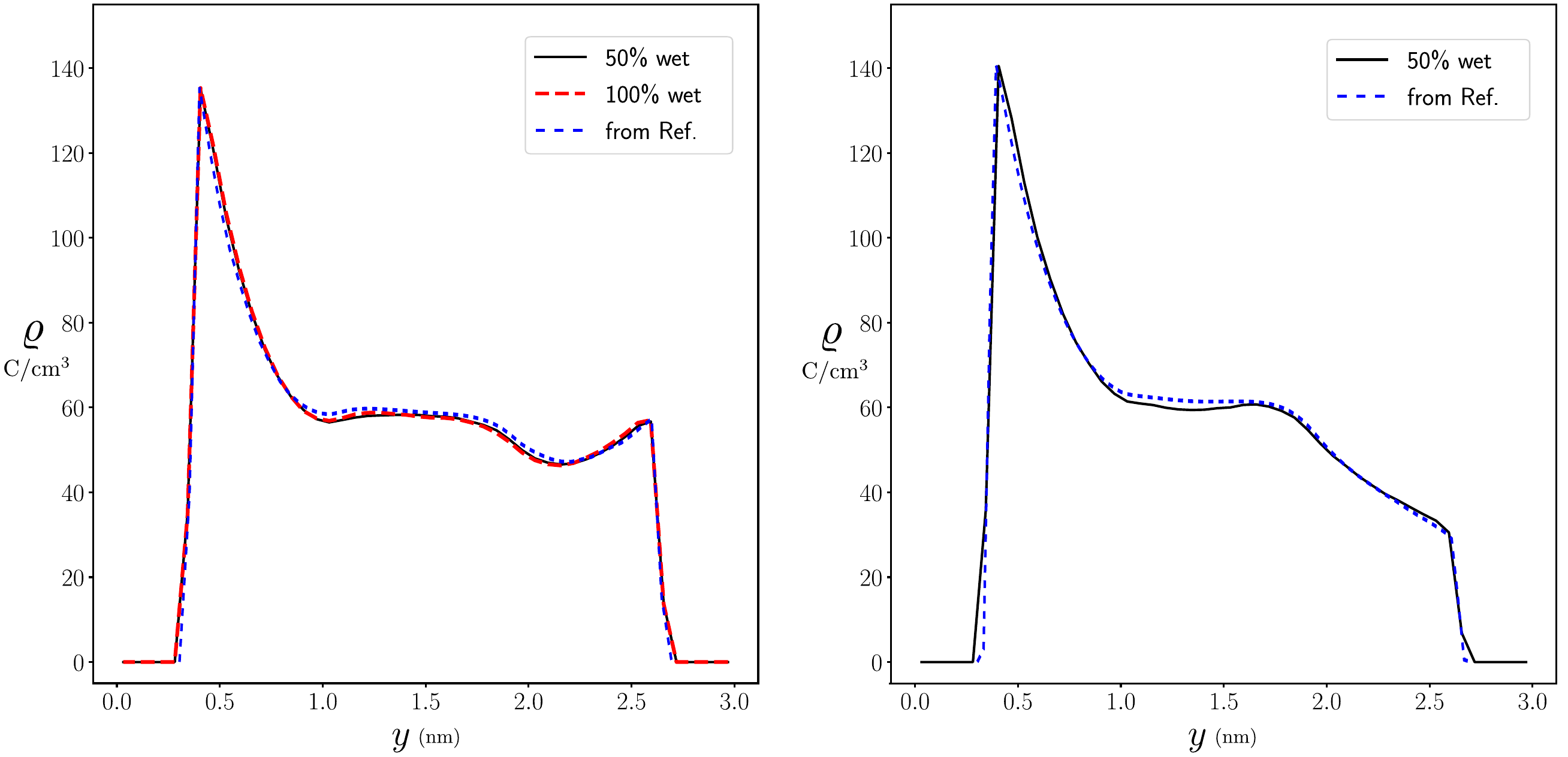}
 \caption{\textcolor{black}{Comparison of the cation density from DISCOS} to Ref.~\citenum{jing2015ionic} for a 1:1 electrolyte (left) and for a 1:2 electrolyte (right). The solid black lines are the DISCOS results using a 50\% wet simulations, the red dashed line indicates a 100\% wet simulation. The blue dotted lines show the OZ/AHNC results from Ref.~\citenum{jing2015ionic}. Note that slight discrepancies between the DISCOS results and those from Ref.~\citenum{jing2015ionic} may be attributed to the difference in inter-particle potentials.}
\label{fig:static}
\end{figure}

The left panel of Fig.~\ref{fig:static}  shows the cation charge density  
in a channel with a 1:1 electrolyte (130 cations and anions, each, with $q_+ = -q_- = 1.6\times 10^{-19} \text{C}$). Each DISCOS simulation was started from a random particle configuration and run for $10^5$ time steps to reach equilibrium. This was then time averaged for at least $10^{6}$ time steps to obtain the results in Fig.~\ref{fig:static}. The right panel is for a 1:2 electrolyte, where $q_-=-3.2\times 10^{-19}$C
and the number of anions has been halved to 65. In both cases good agreement is observed, confirming the accuracy of the DISCOS method in reproducing confined equilibrium configurations.

\subsection{Stochastic drift}\label{sec:result_stochdrift}

As discussed above, for systems at equilibrium the spatial distribution of particles is independent of particle mobility. The particle equation of motion, Eq.~(\ref{eq:ions3}), is dependent on both particle mobility and its divergence, which specifies the stochastic drift. The stochastic drift has three terms:  the wet term given by Eq.~(\ref{eqn:thermforce}), the additional wet term captured by the time stepping scheme given by Eqs.~(\ref{fixman}) and the dry term needed to represent boundary effects. Although the equilibrium result is independent of mobility, obtaining the correct result still depends on correctly calculating the divergence of whatever mobility is being used. In Ref.~\citenum{delong2014brownian} the effect of the wet stochastic drift was demonstrated by measuring the equilibrium distribution in a periodic domain. Although the divergence of the mobility would appear to be zero in this case, there is a small translational variance in the mobility of the Peskin kernels; it was shown that not accounting for this produced a spurious non-uniformity in the distribution of particles. This is demonstrated much more starkly in a confined channel due to the large spatial variation in mobility. 

\begin{figure}[h!]
  \centering
    \includegraphics[width=0.5\textwidth]{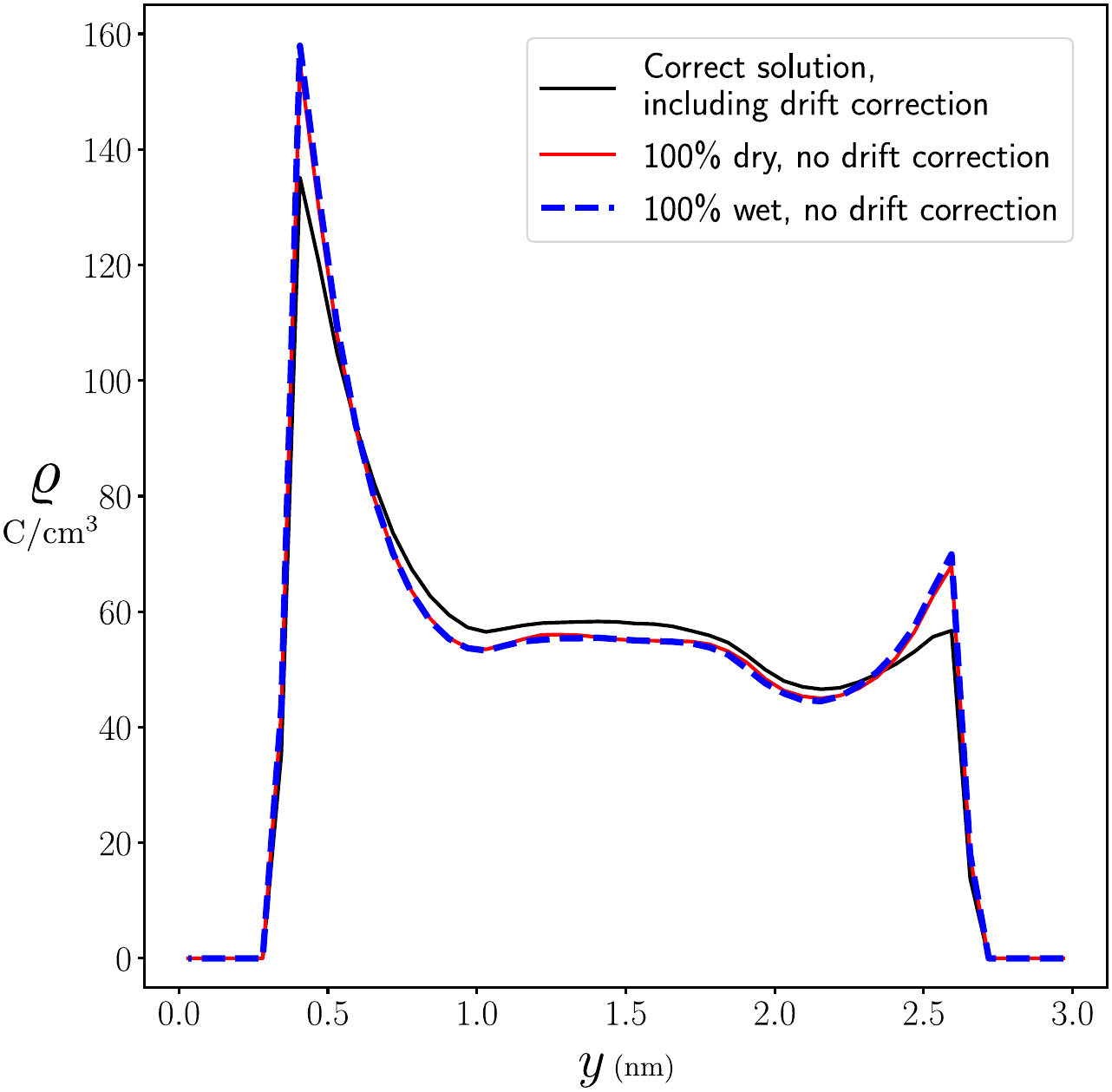}
 \caption{Comparison of error induced by turning off stochastic drift correction.
 Black solid line: \textcolor{black}{Benchmark solution (independent of wet percentage), including stochastic drift correction, from Section~\ref{sec:result_eq}.
 Blue dashed line: 100\% wet simulation without stochastic drift correction.} This is achieved by omitting the RFD term and using a single stepping method in place of the mid-point method; see Sec. \ref{subsec:TemporalAlgorithm}. 
 Red solid line: 100\% dry simulation without stochastic drift correction. This is achieved by omitting the divergence of mobility term in the particle equation of motion; see Eq.~(\ref{eq:ions3}). }
\label{fig:drift}
\end{figure}

In Fig.~\ref{fig:drift}, we have taken the 100\% wet simulation from Fig.~\ref{fig:static} and re-run it with the stochastic drift correction turned off. This is achieved by omitting the $\forcedensity^\text{th}$ term from Eq.~(\ref{stokes_temporal}) and employing a simple single stage time stepping method in place of the midpoint algorithm described in Section~\ref{subsec:TemporalAlgorithm}. Additionally we have run a 0\% wet simulation, where the drift correction has been turned off by omitting the divergence of the diffusion tensor from Eq.~(\ref{eq:ions3}). The correct result (drift correction on) is shown for comparison. In both the 100\% wet and 100\% dry cases, neglecting the drift correction produces an identical result, i.e., they both differ from the correct solution by the same amount. This indicates that the two disparate methods (RFD plus midpoint time stepping for the wet simulation, direct calculation for the dry) predict identical corrections for the stochastic drift, allowing us to validate them against each other.

\section{Electrokinetic Flow Simulations}\label{sec:electrokinetic}

This section presents results from DISCOS simulations of several electrokinetic flows, \textcolor{black}{and examines how they differ from theory and other numerical methods.} Sec.~\ref{sec:result_eo} we examine the non-equillibrium case of electro-osmotic flow driven by an electric field. 
Comparisons are made to theory, previous MD simulations, and results from a deterministic continuum code. In Sec.~\ref{sec:result_iceo} we examine the complex flow patterns associated with induced charge electro-osmotic flow, comparing DISCOS results with theory and a deterministic continuum code.

\subsection{Electroosmotic channel flow}\label{sec:result_eo}
Electrolyte solutions are typically electroneutral in the bulk, however the presence of a surface charge leads to the formation of a diffuse double layer~\cite{newman2012electrochemical}. In the presence of an external electric field this double layer can result in an electrokinetic flow, the most common being electrophoretic and electroosmotic flows. Here we consider the latter, using the same parallel wall geometry as in the previous section to simulate the fluid motion in an electrolyte due to a constant, uniform electric field parallel to the walls.

\subsubsection{Theory}\label{eotheory}

The traditional model for the double layer assumes that the ion concentrations obey a Boltzmann distribution which, combined with the Poisson equation, gives the Poisson-Boltzmann equation
\begin{equation}
\gradspatial^2 \phi = -\frac{1}{\epsilon} \sum_{j=1}^{N_s} \chargedensity_j^0
\exp( - \charge_j (\phi - \phi^0) / \kb T )
\end{equation}
where the subscript $j = 1,...,N_s$ indicates the species and $\chargedensity_j^0 = \charge_j c_j^0$ presents the charge density of species $j$ where $c_j^0$ is the species concentration.
The superscript ``0'' indicates a reference value, which in our
geometry is the center of the channel ($y=L/2)$. In the following sections we consider two regimes \cite{andelman1995electrostatic}, characterized by two length scales: the Debye length and the Gouy-Chapman length, given respectively by
\begin{equation}
\Dlength = \left( \sum_{j=1}^{N_s} \frac{q_j \chargedensity_j^0 }{\epsilon \kb T} \right)^{-1/2}
\mathrm{and}
\qquad  \GClength = \frac{2 \kb T \epsilon}{e |\varsigma|},
  \label{lengths}
\end{equation}
where $e$ is the elementary charge. We also consider two potentials, the electrokinetic potential and thermal potential, given by
\begin{equation}
\phi_\zeta = \Dlength \varsigma /\epsilon \qquad
\mathrm{and}\qquad
\phi_T = \kb T/e.
  \label{potentials}
\end{equation}
As discussed further below, two separate ion concentrations and surface charge densities are considered, where in all cases $\Dlength \ll L$, but where $\GClength$ may be smaller or larger than $\Dlength$.

\textcolor{black}{When $\GClength \gg \Dlength$,} we are in the Debye-H{\"u}ckel regime \cite{debye1923theory,robinson2012electrolyte}. 
Since $\GClength/\Dlength = 2\phi_T/|\phi_\zeta|$
the potential difference in the channel is small compared with the thermal potential and we can linearize the Poisson-Boltzmann equation,
\begin{equation}
    \partial_y^2 \phi \approx -\frac{\chargedensity^0}{\epsilon}
    + \frac{\phi - \phi^0}{\Dlength^2},   \label{debyepotential}
\end{equation}
where $\chargedensity^0 = \sum_{j=1}^{N_s} \chargedensity_j^0$. We now solve Eq.~(\ref{debyepotential}) on $(0,L)$ with 
 $-\epsilon\, \partial_y \phi |_{y=0} = \varsigma $ and 
$-\epsilon\, \partial_y \phi |_{y=L} = -\varsigma $,
to obtain
\begin{equation}
    \phi(y)=\phi^0 + \phi_\zeta \left[1 - \cosh \left( \frac{L-2y}{2 \Dlength} \right) \right]  \mathrm{csch} \left( \frac{L}{2 \Dlength } \right). 
    \label{eq:PhiWalls}
\end{equation}


The fluid velocity can now be computed from the deterministic Stokes equation,
\begin{equation}
    -\eta \partial_y^2 v_x = -\epsilon\, E_x^\mathrm{ext} \partial_y^2 \phi ,
\end{equation}
where we have taken the applied field to be in the $x$ direction.
Solving this equation using no-slip conditions at $y=0$ and $y=L$ then gives 
\begin{equation}
    v_x(y) = -\frac{\epsilon \,\phi_\zeta E_x^\mathrm{ext}}{\eta} \left[\coth\left(\frac{L}{2 \Dlength}\right) - \mathrm{cosh} \left( \frac{L-2y}{2 \Dlength} \right)  \mathrm{csch} \left( \frac{L}{2 \Dlength } \right) \right].   \label{fluidvel1}
\end{equation}
Note that when $\GClength < \Dlength$, we are in an `intermediate' regime \cite{andelman1995electrostatic} where analytic approximation is difficult. In this case we must resort to comparison with other numerical methods.

\subsubsection{Numerical simulations}\label{EO_numeric}

In this section we perform three analyses. In the first, we compare DISCOS to high molarity MD simulations performed in Ref.~\citenum{rezaei2015surface}. Second, we compare DISCOS simulations of moderate molarity to a deterministic continuum code and the above theory. Finally, using the same configuration, we examine the effect of altering the wet/dry diffusion ratio. 

For each simulation we use the general geometry shown in Fig.~\ref{fig:geometry} with $L=P_x=P_z=6\text{nm}$.
For the first analysis we have set our simulation parameters to match the MD parameters as closely as possible, noting that this is not always possible to do exactly due to the continuum components of DISCOS. The details of this are discussed in Appendix~\ref{appdx1}.

All simulations are performed with an imbalance of anions and cations in the channel combined with inhomogeneous Neumann condition \textcolor{black}{representing a constant surface charge (the same on both walls),} so that the overall electrostatic system is charge neutral. In all simulations we have employed the 4-point Peskin kernel for electrostatic calculations. For the hydrodynamic calculation we make use of the exponential of a semicircle (ESC) kernel \cite{FINUFFT_Barnett}. This generalized kernel has several free parameters that allow small adjustments to the hydrodynamic radius. Using this approach we are able to match the desired anion and cation diffusion coefficients to within 2\% using a grid of $56^3$ cells. An 85\% wet simulation can be performed with a $48^3$ grid, and 50\% with $28^3$. The ESC kernel is discussed further in Appendix~\ref{appdx2}. Note that in all cases we have used the same grid for the electrostatic solution.

\begin{figure}[h!]
  \centering
    \includegraphics[width=0.95\textwidth]{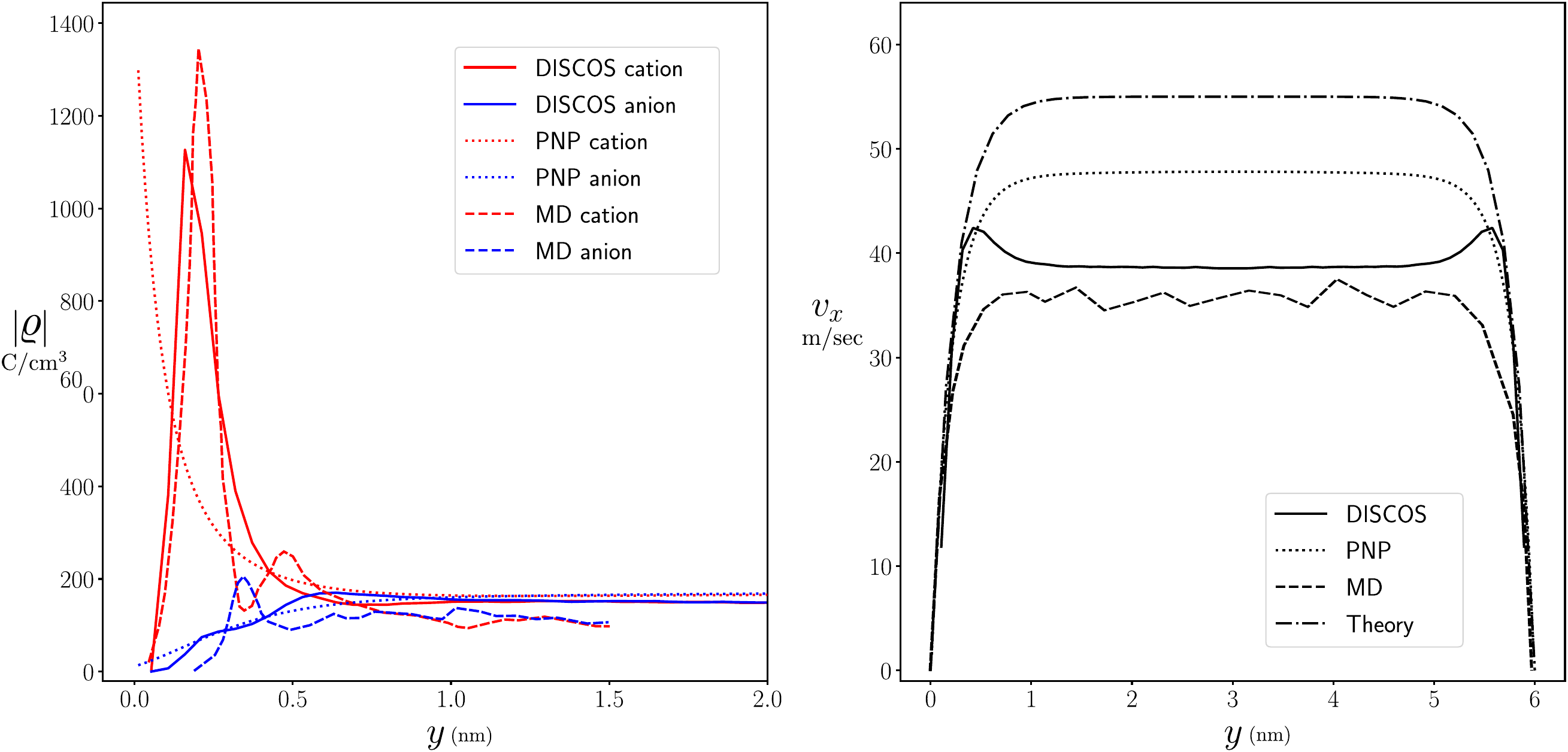}
 \caption{Left: a comparison of the absolute charge distributions between \textcolor{black}{DISCOS, PNP simulations,} and the MD results for \textcolor{black}{high molarity from Ref.~\citenum{rezaei2015surface}.} The MD simulations show a complex three layered structure. DISCOS captures two of these layers, while the continuum simulation displays a single layer structure. Right: The fluid velocity profiles resulting from DISCOS, continuum, and MD simulations, compared with Debye theory. \textcolor{black}{For the DISCOS simulations the velocity field is sampled after step three in the algorithm given in Section.~\ref{subsec:TemporalAlgorithm}.} Note that, unlike the following examples, the statistical error in the DISCOS results is negligible due to the high ion concentration.}
\label{fig:EO1}
\end{figure}

\begin{center}
\emph{High molarity}
\end{center}
In Fig.~\ref{fig:EO1} we show the results of a simulation of with a total ion concentration of 3.52M, corresponding to 269 $\mathrm{Na}^+$ ions and 189 $\mathrm{Cl}^-$ ions, leading to a surface charge density of $\varsigma = -1.45\times10^{-5} \text{C}/\text{cm}^2$ so that the overall system is electroneutral; this corresponds to one of the MD cases examined in Ref.~\citenum{rezaei2015surface}, which is also shown. The simulation was run for $10^5$ time steps to reach steady state, and then time averaged for a further $4\times10^4$ steps to obtain the results shown here. For contrast we have also included results obtained using the theory above, \textcolor{black}{and with a continuum code for simulating electrolytes solving\footnote{\textcolor{black}{In fact, the same underlying Stokes and Poisson solvers are used by both codes.}} the non-linear Poisson-Nersk-Planck (PNP) equations \cite{peraud2016low}. Although one can include fluctuations in this formulation,} they are not valid at this scale because of the small number of ions per computational cell.
We therefore perform the continuum simulations deterministically, i.e., without thermal fluctuations. The relevant length scales are $\Dlength/L=0.038$ and $\GClength/L=0.028$, as this is beyond the range of validity for the Debye theory discussed above, we expect a discrepancy between the DISCOS and MD results when compared to the theory of Sec.~\ref{eotheory}.

The increased importance of boundary effects is clearly visible in the left plot of Fig.~\ref{fig:EO1}, with the charge distribution simulated by DISCOS being closer to the MD result than the continuum. The charge distribution from the MD simulations shows a clear three-layered structure: closest to wall is a large cation peak, followed by a smaller layer of anions, which is followed in turn by an additional layer of cations. This is discussed in detail in Ref.~\citenum{rezaei2015surface}. The DISCOS simulation captures the inner two of these layers, although they are somewhat more diffuse than those in the MD simulation. The effect of the anion layer in the DISCOS simulation is visible in the velocity profile shown in the right plot: the majority of the flow is generated by the motion of the cation layer, but the opposite motion of the anion layer reduces the fluid velocity somewhat in the center of the channel. 

In Ref.~\citenum{rezaei2015surface}, it is shown that increasing the magnitude of the surface charge can cause this effect to completely reverse the flow. We note that in this regime of extremely high concentrations and surface charge densities the bulk properties of the flow are extremely sensitive to the surface parameters, and, when using the surface configuration shown in Fig.~\ref{fig:EOdiag}, DISCOS does not achieve the same flow reversal effect. 

\textcolor{black}{As touched on in Appendix~\ref{appdx1}, the most likely cause of this discrepancy is the continuous nature of the hydrodynamic and electrostatic boundary conditions in the DISCOS simulations, versus the discrete array of particles used in the MD simulations; see Fig.~\ref{fig:EOdiag} and Ref.~\citenum{rezaei2015surface}. An illustration of the importance of this effect is given in Ref.~\citenum{freund2002electro}. In that article, similarly to Ref.~\citenum{rezaei2015surface}, electro-osmosis in a channel with walls comprised of a silicon lattice is examined. Changing the configuration of surface charge from being evenly distributed among all the surface atoms (as in Ref.~\citenum{rezaei2015surface}), to being concentrated in a small subset of the surface atoms, produced a 25\% change in peak flow velocity. An interesting avenue of investigation for DISCOS would be to construct the walls out of particles in a similar manner to MD, allowing a discrete representation of the surface charge.}


\begin{figure}[h!]
  \centering
    \includegraphics[width=0.95\textwidth]{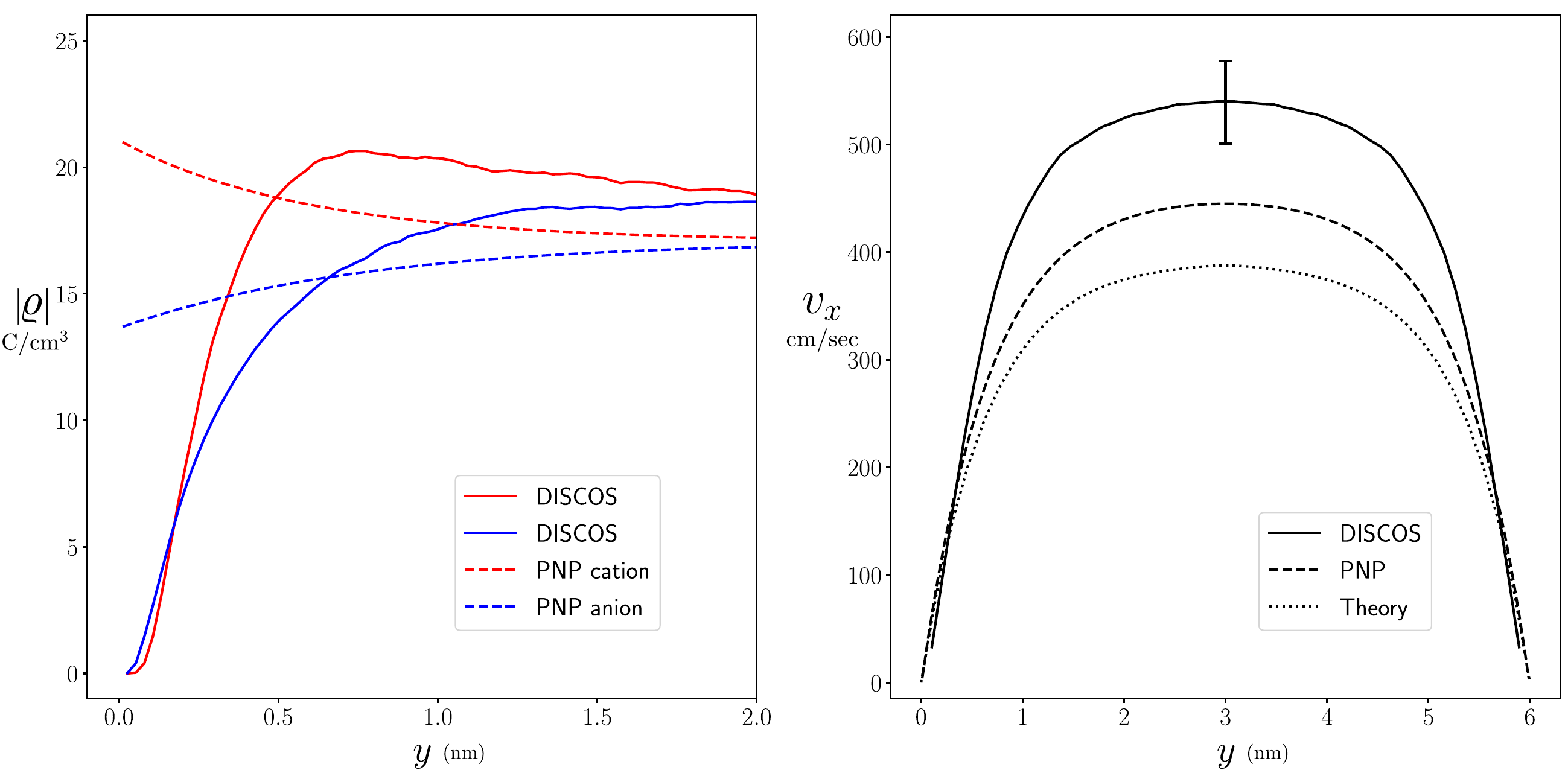}
 \caption{Electro-osmosis simulations. Left: Comparison of charge distributions near the wall resulting from DISCOS and the PNP code \cite{peraud2016low}\textcolor{black}{ at moderate molarity.} As above, the DISCOS result differs substantially from the continuum prediction due to short range inter-particle interactions. 
 Right: Resulting flow velocity. Due to the differing charge profiles, the DISCOS simulation produces a flow velocity substantially higher than Debye theory, and simulations using a continuum code. \textcolor{black}{Note that the error bar denotes $\pm$ two standard errors.}}
\label{fig:EO0}
\end{figure}

\begin{center}
\emph{Moderate molarity}
\end{center}
In Fig.~\ref{fig:EO0} we show the results of a simulation with a total ion concentration of 0.354M, corresponding to 24 $\mathrm{Na}^+$ ions and 22 $\mathrm{Cl}^-$ ions, which leads to a surface charge density of $\varsigma = -4.44\times10^{-7} \text{C}/\text{cm}^2$. Again, the simulation was run to steady state for $10^5$ time
steps, and in this case averaged for $1.3\times10^6$ steps. 
Comparison is made with the theory of Sec.~\ref{eotheory} and with the \textcolor{black}{non-linear PNP code.} Here, using the center channel charge density we find $\Dlength/L=0.094$ and $\GClength/L = 1.1$. As neither of these approaches incorporate short range effects, in order to make a closer comparison we have modified the parameters of the DISCOS simulation to remove the short range boundary condition described in Appendix~\ref{appdx1} (all other parameters remain the same). Instead, the centroids of ions encountering the boundary are specularly reflected.

As can be seen in the charge distribution profiles, even in the absence of short range
effects from the boundary, packing effects from particle interactions \textcolor{black}{(including electrostatic interactions with image image charges) not accounted for in the PNP equations} cause very different charge profiles between DISCOS and the continuum simulations. Consequently, DISCOS yields a noticeably higher flow velocity than the continuum code and the theoretical prediction. Even though the this simulation is performed with $\Dlength$ and $\GClength$ values where we might expect closer agreement with theory, the proportional difference in the results is similar to the case shown above; this highlights the importance of short range particle interactions (not included in the theoretical or PNP results) when performing simulations at scales where the channel width, $L$, begins to become comparable to the short range interaction length, $\sigma$. \textcolor{black}{Note that it is not possible to turn off steric interactions between particles in discrete ion simulations.}

\begin{figure}[h!]
  \centering
    \includegraphics[width=0.95\textwidth]{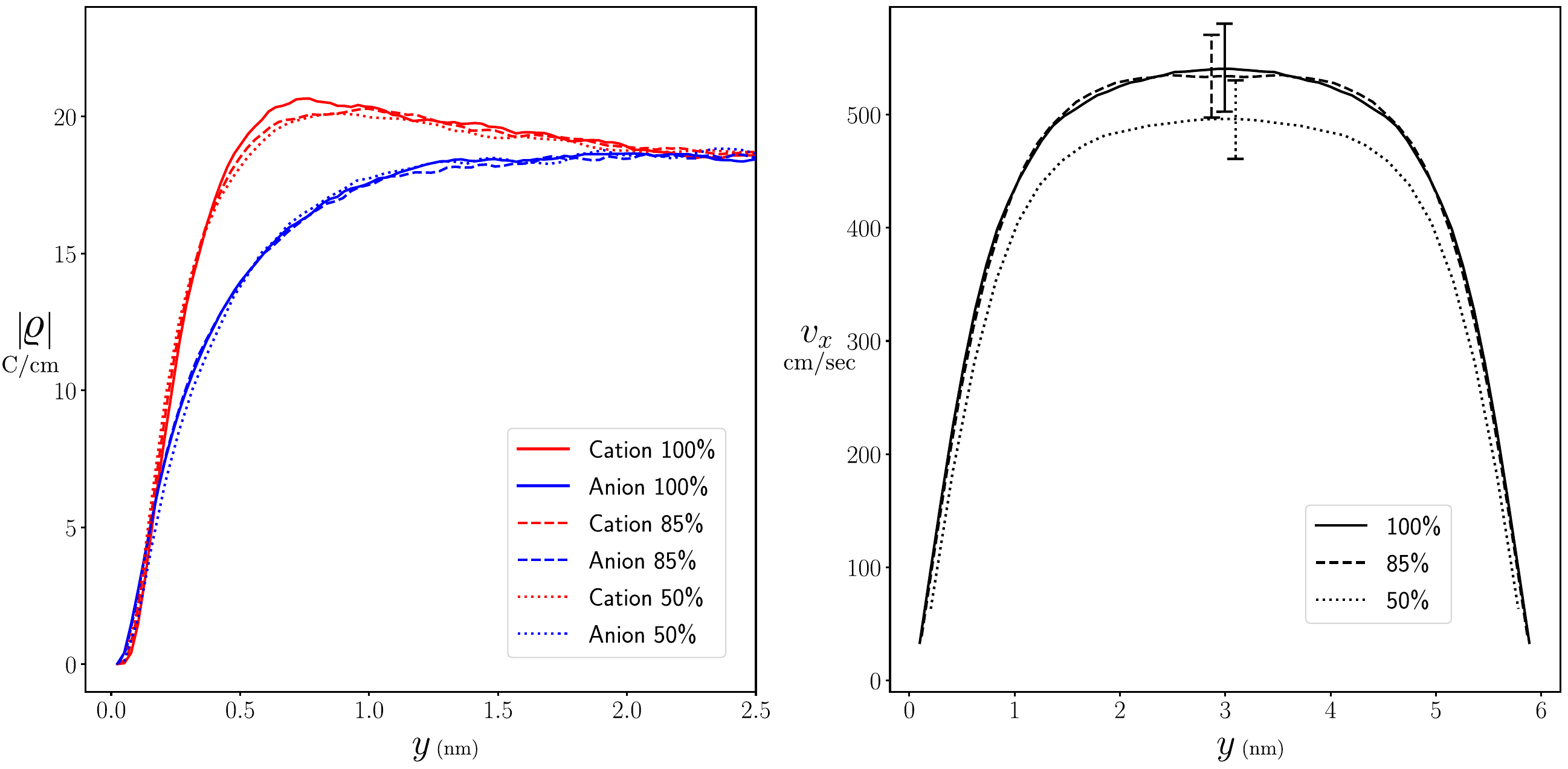}
 \caption{Electro-osmosis simulations at different wet percentages. Left: Comparison of charge distributions near the wall. Right: Comparison of velocity profiles. \textcolor{black}{Using a moderate amount of dry diffusion, 15\%, runs roughly twice as fast while having a negligible impact on the results. Using 50\% dry diffusion the error in peak velocity is about 8\%, with an eight fold compulational speedup. Note that the error bars denote $\pm$ two standard errors.}}
\label{fig:EO2}
\end{figure}

\begin{center}
\emph{Dry diffusion}
\end{center}
\textcolor{black}{In Fig.~\ref{fig:EO2} we examine the effect of dry diffusion. The moderate molarity simulation has been re-run with a grid of $48^3$ cells, corresponding to 85\% wet, and $28^3$ cells, corresponding to 50\% wet. The effect of the walls on the dry mobility has been incorporated following the approach of sections \ref{sec:drymob} and \ref{sec:result_drymob}. The 85\% cases shows a negligible difference to the 100\% wet simulation, in both the charge distribution and velocity profile. In agreement with Ref.~\cite{ladiges2020discrete}, at moderate concentrations some degree of dry diffusion can be used with little impact on accuracy, allowing a large computational speedup. This approach is especially effective for channel electro-osmosis as the dominant effect on particle mobility is drag from the boundary, which has been fully accounted for as shown in Sec.~\ref{sec:result_drymob}. The 50\% wet simulation is approximately 8 times faster, whilst giving an error in peak velocity of approximately 8\%. This suggests that approximate answers can be quickly obtained using larger amounts of dry diffusion, again in agreement with Ref.~\cite{ladiges2020discrete}.}

\subsection{Induced charge electroosmosis}\label{sec:result_iceo}
\begin{figure}[h!]
  \centering
    \includegraphics[width=0.95\textwidth]{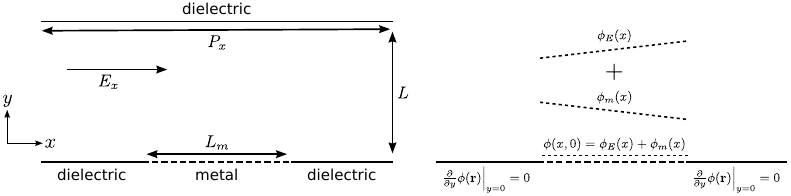}
 \caption{Left: ICEO geometry. Right: Illustration of electrostatic boundary condition. The solid lines represent dielectric regions of the boundary where a homogeneous Neumann condition is applied when solving Eq.~(\ref{eq:poisson}), the dashed line represents the metal strip where a fixed potential is applied. The external electric field has a potential represented by $\phi_E$. The fixed potential boundary condition used when solving Eq.~(\ref{eq:poisson}) is $\phi_m$. This is set such that the total potential is constant on the metal strip.
}
\label{fig:EOgeometry}
\end{figure}

As a final demonstration that DISCOS captures microscopic dynamics of an electrolyte, we perform simulations of a more complicated 
electrokinetic flow:
induced-charge electro-osmosis (ICEO) \cite{squires2004induced}. One typical realization of ICEO is a channel with walls composed of dielectric material, with a 
metal strip placed on one of the boundaries as illustrated in Fig.~\ref{fig:EOgeometry}.

An external electric field is applied in the $x$ direction, tangential to the surface. This induces the ions to form an electrical double layer on the metal strip with a charge density gradient, i.e., with positive ions gathering at the right side of the metal plate and negative ions at the left side. Consistent with Eq.~(\ref{fluidvel1}), the fluid velocity scales with the electrokinetic potential, $\phi_\zeta$, and the external field $E_x^\mathrm{ext}$.
Note that the electrokinetic potential represents the drop in electric potential across the Debye layer \cite{IUPAC1997compendium}, and 
scales as $\phi_\zeta \sim E_x^\mathrm{ext}L_m$ (where $L_m$ is the length of the metal strip) due to the surface charge gradient along the metal strip. As a result, the characteristic fluid velocity scales quadratically with the applied field in the small $\phi_\zeta$ regime. Overall, the fluid is pushed in opposite directions toward the center of the metal plate because $\phi_\zeta$ is positive on one side of the metal strip (where cations accumulate) and negative on the other side (where anions accumulate), forming a counter-rotating vortex pair. 
\begin{figure}[h!]
  \centering
    \includegraphics[width=0.95\textwidth]{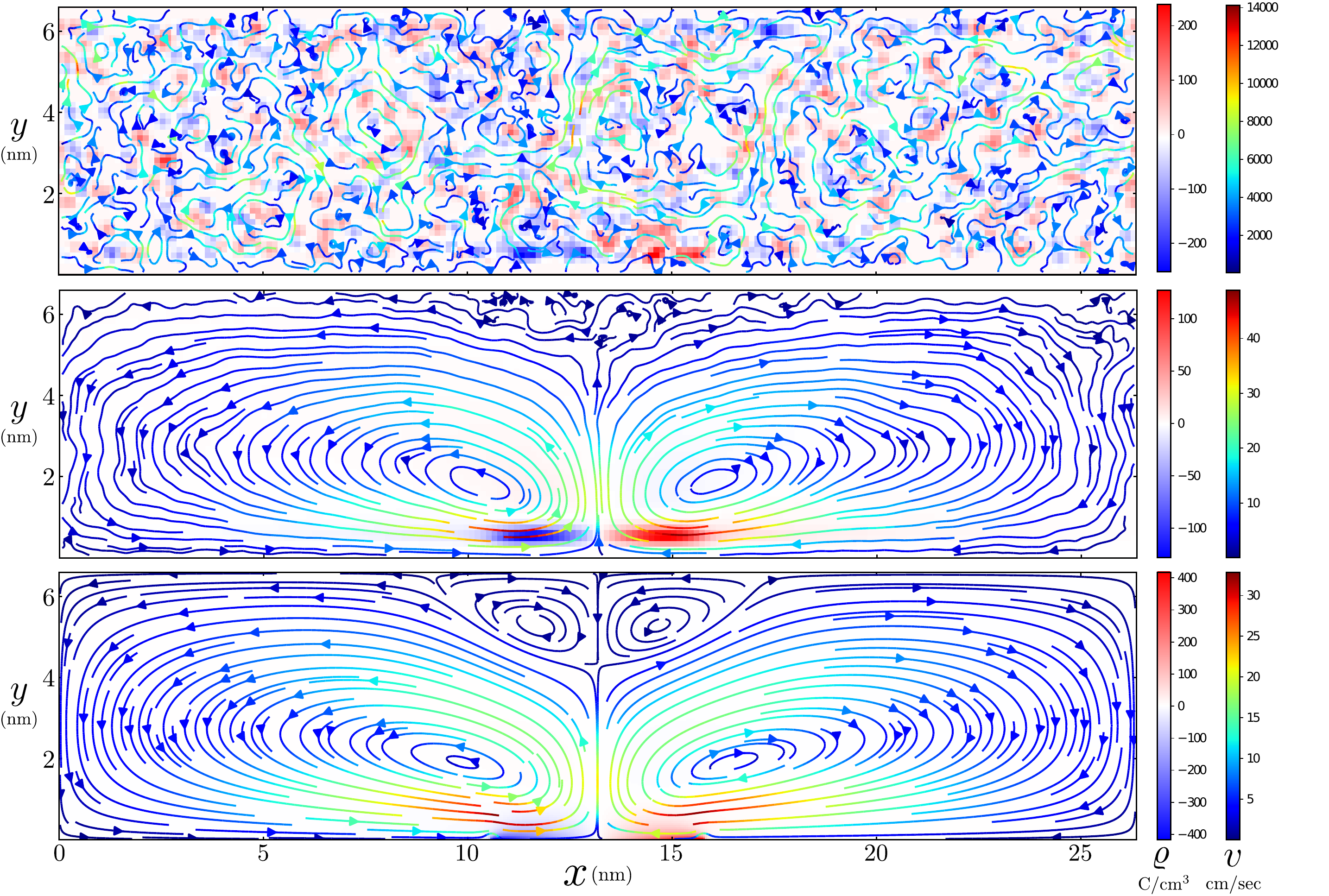}
    \caption{Contour plots of charge density (left color bars) and flow field speed (right color bars) of ICEO at $\mathbf{E}=[-10^8,0,0]$~V/m. Top: \textcolor{black}{A snapshot of fluid velocity in a DISCOS simulation,} averaged in $z$-direction. Middle: DISCOS simulation result, averaged in $z$-direction, then time-averaged after initial 40,000-step transient period and ensemble-averaged across 20 independent runs. Each simulation has run for a timescale on the order of $40$~ns ($O(400,000)$ steps), so overall averaged over 8 million steps. Bottom: deterministic continuum simulation using the PNP equations.}
    \label{fig:iceo_flow}
\end{figure}

We conduct our ICEO simulation in a channel with $L=6.59$~nm, and $P_x=P_z=26.36\text{nm}$. The metal strip has length $L_m=5.272\text{nm}$ and is centered on the lower boundary; see Fig.~\ref{fig:EOgeometry}. The electrolyte is a 0.15M 1:1 solution at temperature $T=300\,\mathrm{K}$ with cation and anion charges $q_{+}=-q_{-}=1.6 \times 10^{-19} \, \textrm{C}$, and diffusion coefficients $D^\text{tot}_A = D^\text{tot}_C = 1.89\times 10^{-5}\text{cm}^2\text{s}^{-1}$. The solvent is water with viscosity $\eta=0.009\,\mathrm{g/(cm \cdot s)}$ and relative permittivity $\epsilon_r=80$. \textcolor{black}{Using a hydrodynamic grid of $192\times48\times192$ cells with the four point Peskin kernel results in the simulation that is 75\%} \textcolor{black}{wet; the same grid and kernel is used for the Poisson solve}, \textcolor{black}{as mentioned previously, the P3M radius is set to match the radius of support of the kernel.} Homogeneous Neumann conditions used for the dielectric parts of the boundary, and fixed potential Dirichlet potential for the part representing the metal strip. This is set such that the potential on the surface of the metal is constant; see Fig.~\ref{fig:EO1}. In this study we present results for an applied field of $E_x=-10^8\,\text{V/m}$. For short range interactions with the walls we have used a Lennard-Jones 9-3 potential, with $\xi=7.95 \times 10^{-21}\text{J}$ and $\sigma=0.426\text{nm}$. The Lennard-Jones potential is also used for interparticle short range interactions, with $\xi=8.16 \times 10^{-22}\text{J}$ and $\sigma=0.442\text{nm}$. \textcolor{black}{Note that near the metal-dielectric transition points the image charge construction is approximate, since the spreading and interpolation operators might span into regions with a different boundary condition. However for these simulation parameters $\sigma$ is slightly larger than the radius of support of the kernel, $2\Delta r= 4.11\text{nm}$. This prevents ions from approaching too close to the boundary and minimizes the error from the image charge construction.}

\begin{figure}[h!]
    \centering
    \includegraphics[width=0.95\textwidth]{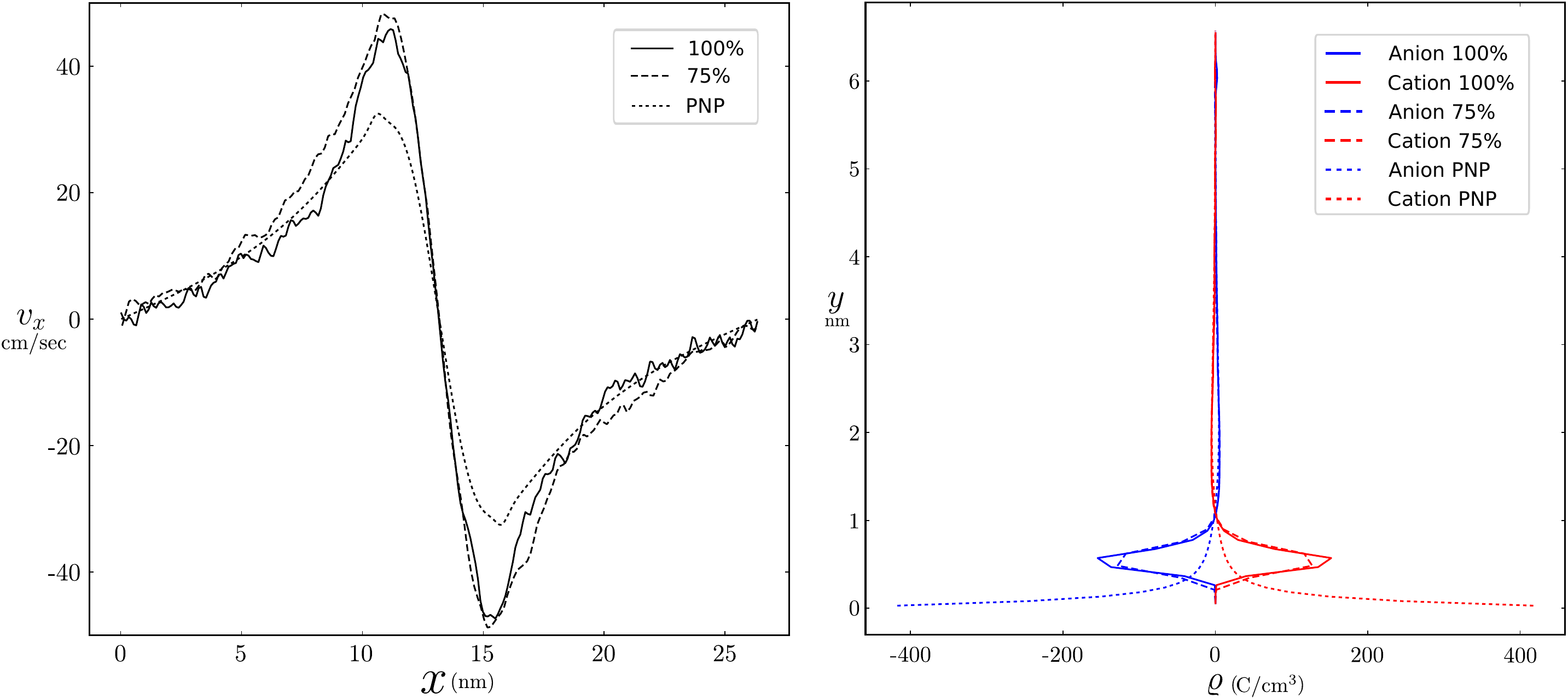}
    \caption{\textcolor{black}{Left: Comparison of DISCOS and deterministic PNP simulations x-velocity near the surface of the bottom plate ($y=0.515\,\mathrm{nm}$). Right: Charge density profiles in y-direction at the $x$ locations where the absolute charge density is greatest ($x = 11.64 \; \mathrm{and} \; 14.72 \, \mathrm{nm}$). 
    For comparison we have included a 100\% wet DISCOS simulation, with $192\times48\times192$ cells. Both the 75\% and 100\% wet simulations predict the same peak velocity.}}
    \label{fig:iceo_ux}
\end{figure}

The simulation results are shown in Figs.~\ref{fig:iceo_flow} and \ref{fig:iceo_ux}, where once again we make a comparison to deterministic continuum results that solve Poisson-Nernst-Planck (PNP) equations using the method outlined in Ref.~\citenum{peraud2016low}. To obtain the DISCOS result, the simulations are averaged in the $z$-direction, then time-averaged over 400,000 steps for each simulation (after an initial equilibration time of 40,000 steps), and finally ensemble-averaged across 20 independent simulations. For comparison the spanwise (z) average from a single step is also shown. The expected counter rotating vorticies are clearly visible in the time averaged data, as is the polarization of charge on the surface of the metal plate. Once again differences are observed between the DISCOS and PNP results, likely arising from the lack of fluctuations and differences in steric effects in the PNP simulations. As in the previous section, this is most notable in the charge distribution close to the surface, shown in Fig.~\ref{fig:iceo_ux}. \textcolor{black}{The PNP code allows charge to build up on the boundary, while the steric effect in the DISCOS simulations prevents this from occurring. Interestingly, despite the difference in the charge distribution, the location of the peak velocity, indicated by dark red color on streamlines, is roughly the same for both DISCOS and the PNP code, which is probably due to the no-slip boundary condition for the fluid. The PNP code also produces a velocity profile quite close to the DISCOS result in this ICEO example, except for the value of the peak velocity; the peak velocity in the DISCOS simulations is about 50\% higher, a larger discrepancy than that seen the in electro-osmotic flows examined in Fig.~\ref{fig:EO0} of the previous section.}

This configuration yields a Debye length $\Dlength=1.15 \,\mathrm{nm}$, and we find $\phi_\zeta=122.5 \,\mathrm{mV}$. These two parameters give a dimensionless Dukhin number \cite{lyklema2000fundamentals} that indicates the effect of surface conduction on the flow,
\begin{equation}
    \mathrm{Du} = \frac{\Dlength}{L_m} \exp \left(\left|\frac{\phi_\zeta}{2 \phi_T}\right|\right)
    = \frac{\Dlength}{L_m} e^{\Dlength/\GClength}~.
\end{equation}\label{eqn:du}
For our ICEO simulation, $\mathrm{Du}=2.32$, which is $O(1)$, suggesting that we are in the large electrokinetic potential regime where the surface conduction becomes significant. As such, the simple theoretical scaling of the electrokinetic potential ($\phi_\zeta \sim E_x$) and flow velocity ($v_x \sim \phi_\zeta E_x \sim E_x^2$) with the electric field are expected to break down \cite{schnitzer2012induced}. We will examine this effect in a future paper.

\section{Summary and Conclusions}\label{sec:conclusion}

In this paper, we have extended the DISCOS method for electrolytic flows to account for the presence of confining boundaries. In Section~\ref{sec:Results}, the ion mobility and equilibrium distributions were validated against theory and existing numerical results. In Section~\ref{sec:electrokinetic}, we examined the use of DISCOS for two electro-kinetic flows: electro-osmosis and ICEO. In all cases DISCOS produced the expected features of the flow, with results similar to those obtained using alternative methods. The principle difference between DISCOS, continuum simulations, and theory arises from the presence of steric effects in DISCOS.

The differences noted between molecular dynamics simulations and those obtained by DISCOS suggest a further avenue of investigation. For systems with increasingly narrow channels, and increasingly high surface charges, the details of the short range interation of the ions and the walls becomes increasingly important. As noted in Section \ref{sec:result_eo}, the continuum description of the hydrodynamic and electrostatic boundary fails to account for the discrete nature of the charged particles forming the wall. An alternative approach would be to use DISCOS particles to form the channel walls. This would produce electrostatic and hydrodynamic behavior closer to that of the MD simulations. Another possible avenue of development would be to use a hybrid approach where DISCOS was used for the bulk of the flow, and MD for the surface layer; such an approach using the SELM method is examined in Ref.~\citenum{guo2016multi} (see also Refs.~\citenum{DelgadoBuscalioniReview} and \citenum{HybridsBell2020}). As the inclusion of MD comes with a large computational cost, examining the use DISCOS particles to capture as much of the wall dynamics as possible would appear to be beneficial.

Conversely, in future work we would like to be able to model electrokinetic flows at larger scales. Although the DISCOS methodology is considerably more efficient than MD, it would be computationally expensive for micron scale flows. Furthermore, for larger scale flows, the electric double layer where fine-scale resolution is needed is confined to a very small fraction of the domain so the bulk of the problem can modeled with continuum FHD electrolyte code \cite{peraud2016low,donev2019fluctuating2} on a coarser grid. These larger-scale problems could therefore be addressed with a hybrid algorithm that couples DISCOS near boundaries to the continuum FHD electrolyte model for the bulk flow; this has some similarity to the Adaptive Mesh and Algorithm Refinement (AMAR) approach \cite{AMAR_DSMC}. One interesting question that arises from this is how to correctly discretize the fluctuating Stokes equations at the internal boundary where we transition from the coarse grid used by the continuum solver to the finer grid used by DISCOS. \textcolor{black}{For stochastic equations care must be taken to preserve the correct statistical properties} (e.g., fluctuation-dissipation balance). This has been examined in the context of SELM using a finite element method in Ref.~\citenum{plunkett2014spatially}, an equivalent approach would have to be derived for the overdamped finite volume context of DISCOS. Aside from its utility in developing a hybrid approach, this would also enable mesh refinement to be employed in purely DISCOS simulations.

\section*{Acknowledgements}
This work was supported by the U.S.~Department of Energy, Office of Science, Office of Advanced Scientific Computing Research, Applied Mathematics Program under contract No.~DE-AC02-05CH11231.
This research used resources of the National Energy Research Scientific Computing Center, a DOE Office of Science User Facility supported by the Office of Science of the U.S. Department of Energy under Contract No.~DE-AC02-05CH11231. A. Donev was also supported by the NSF under awards DMS-2011544 and CBET-1804940.

\begin{appendix}
\section{Electro-osmosis simulation parameters}\label{appdx1}

For the simulations in Sec.~\ref{EO_numeric} we have chosen to replicate as closely as possible the conditions used in the MD simulations of Ref~.\citenum{rezaei2015surface}. The anions were taken to be $\ClNeg$ (chloride), with $q_\text{A} = -1.6\times 10^{-19}\text{C}$, and the cations $\NaPos$ (sodium), with $q_\text{A} = -q_\text{C}$ (the subscripts A and C refer to anion and cation). The diffusion coefficients of these ions have been measured in water at 300K as $2.03\times 10^{-5}\text{cm}^2\text{s}^{-1}$ and $1.33\times 10^{-5}\text{cm}^2\text{s}^{-1}$ respectively, however Ref.~\citenum{rezaei2015surface} uses the simple point charge (SPC) \cite{berendsen1981interaction} water model. SPC water has a viscosity of $0.004\mathrm{g/(cm \, s)}$ \cite{wu2006flexible}, compared to $0.0085\mathrm{g/(cm \, s)}$ for real water at 300K. We have therefore adjusted the diffusion coefficients to $D^\text{tot}_A = 4.31\times 10^{-5}\text{cm}^2/\text{s}$ and $D^\text{tot}_C = 2.94\times 10^{-5}\text{cm}^2/\text{s}$ via the Einstein relation (Eq.~(\ref{einstein})). The relative permittivity is set to $\epsilon_r=66.29$, again to match the SPC water model \cite{wu2006flexible}. In all cases the flow is driven by an external electric field in the $x$ direction of $5.5\times10^{8}\text{V}/\text{m}$. 

\begin{figure}[h!]
  \centering
    \includegraphics[width=0.95\textwidth]{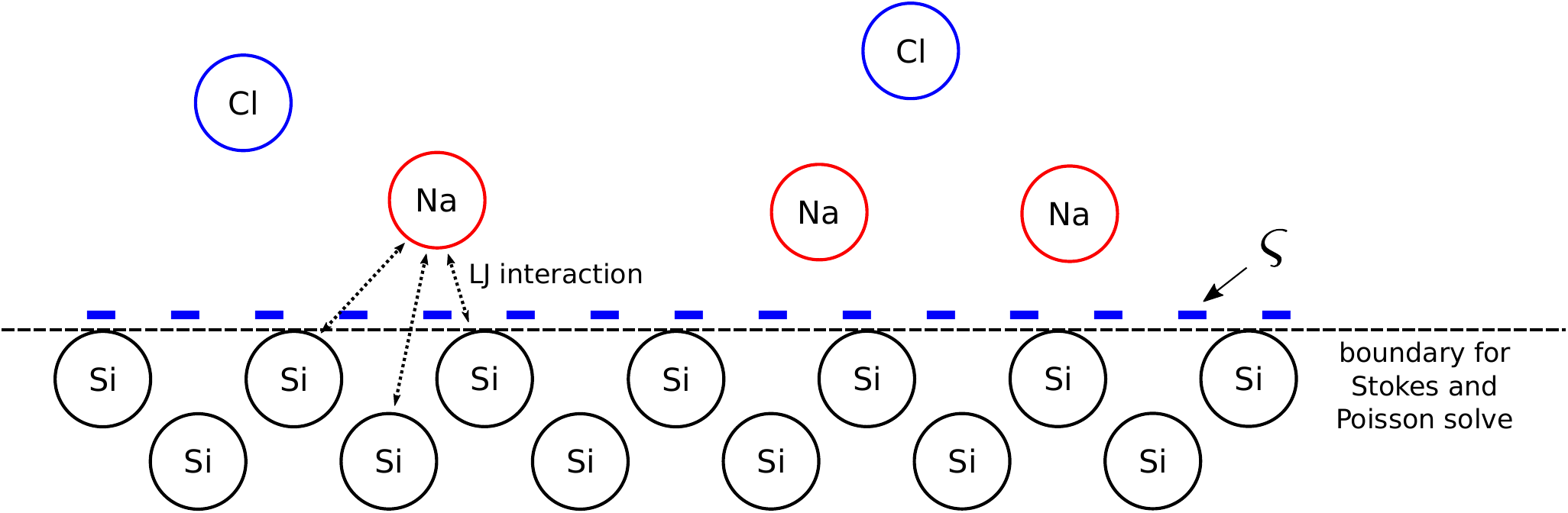}
 \caption{Surface interactions for DISCOS simulations. The LJ interactions have been set to match those from the silicon lattice used in the MD simulations of Ref.~\citenum{rezaei2015surface}. In the MD simulations, each of the inner most layer of silicon atoms is assigned a charge to give the desired surface charge density. In the DISCOS simulations the surface charge is represented as being continuous, via the Neumann boundary condition to Poisson's equation. This contributes to some significant differences between DISCOS and MD simulations for higher surface charge densities. }
\label{fig:EOdiag}
\end{figure}

Again following Ref.~\citenum{rezaei2015surface}, we use a Lennard-Jones potential for short range interactions with the following diameters: $\sigma_\text{AA} = 0.445\text{nm}$, $\sigma_\text{CC} = 0.258\text{nm}$, and $\sigma_\text{AC} = 0.339\text{nm}$. The corresponding magnitudes are given by $\xi_\text{AA} = 7.37\times 10^{-22}\text{J}$, $\xi_\text{CC} = 1.02\times 10^{-22}\text{J}$, and $\xi_\text{AC} = 2.75\times 10^{-22}\text{J}$.
In all cases the cutoff was set to $1.1\text{nm}$. In the referenced MD simulations, the walls are formed using a lattice of four layers of silicon atoms in the 111 orientation. To replicate this as closely as possible, as an ion approaches a wall a Lenard-Jones force is calculated from the arrangement of silicon atoms depicted in Fig.~\ref{fig:EOdiag}. In this case the LJ parameters are $\sigma_\text{AS} = 0.388\text{nm}$, $\sigma_\text{CS} = 0.296\text{nm}$, $\xi_\text{AS} = 2.12\times 10^{-21}\text{J}$, and $\xi_\text{CS} = 3.88\times 10^{-22}\text{J}$, where the S subscript refers to a silicon atom.

Using this approach, the LJ interactions with the boundary are similar to those in the MD simulations, however electrostatic interactions are still with a uniform surface charge; in the MD simulations charge is distributed discretely among surface atoms. Similarly, the hydrodynamic boundary is a smooth surface rather than the `rough' silicon lattice seen in the MD simulations. Both of these effects contribute to some difference between the DISCOS and MD simulations.

\section{Exponential of a semicircle kernels}\label{appdx2}

For spreading and interpolation, in this paper we have made use of three IB kernels. In sections \ref{sec:result_eq}, \ref{sec:result_stochdrift}, and \ref{sec:result_iceo}, the four point Peskin kernel \cite{peskin2002} has been used for cations and anions, for both hydrodynamic and electrostatic calculations. It is defined using
\begin{align}
\delta^\mathrm{Pe}(\zeta_k)= 
\hspace{-1mm}\begin{cases} 
      \frac{\displaystyle 3-2|\zeta_k| + \sqrt{\displaystyle 1+4|\zeta_k| -4|\zeta_k|^{2}} }{\displaystyle 8 \Delta \absspatial}, &  
      \hspace{-2mm} 0 \leq  |\zeta_k|  \leq 1 \vspace{2mm} \\
      \frac{ \displaystyle 5 - 2|\zeta_k| - \sqrt{\displaystyle -7+12|\zeta_k| -4|\zeta_k| ^2}}{{\displaystyle 8 \Delta \absspatial}}, & 
      \hspace{-2mm} 1 <|\zeta_k| \leq 2  \vspace{2mm}\\
       0, &\hspace{-2mm}  2<|\zeta_k| 
   \end{cases},\label{peskin1d}
\end{align}
where $\bs{\zeta} =(\position_i-\spatial)/\Delta \absspatial$, and $\zeta_k$ indicates a single Cartesian component of $\bs{\zeta}$, ($\zeta_x$, $\zeta_y$, or $\zeta_z$). The hydrodynamic kernel is then given by
\begin{align}
\hydrokernel(\bs{\zeta})=\delta^\mathrm{Pe}(\zeta_x)\delta^\mathrm{Pe}(\zeta_y)\delta^\mathrm{Pe}(\zeta_z),\label{eq:pkernel}
\end{align}
with the electrostatic kernel, $\electrokernel$, defined similarly. 

To increase the range of hydrodynamic radii and expand to more realistic electrolytes, DISCOS also uses a recently developed kernel \cite{FINUFFT_Barnett} called the ``exponential of a semicircle" (ESC) kernel. The functional form of this kernel is given by 
\begin{align}
    \delta^\mathrm{ESC}(z; \alpha, \beta) = \frac{1}{\int^{\alpha}_{-\alpha} \exp\left[\beta(\sqrt{1-(\frac{z}{\alpha})^2}-1)\right]dz}
    \begin{cases} 
        \exp\left[\beta(\sqrt{1-(\frac{z}{\alpha})^2}-1)\right], & |\frac{z}{\alpha}| \leq 1 \\
        0, & |\frac{z}{\alpha}| > 1,
    \end{cases}
    \label{eq:eskernel}
\end{align}
where $\alpha=w\Delta r/2$, $w$ is the number of points for spreading or interpolation, and $\beta$ is a tuning parameter that changes the hydrodynamic radius. This is then used in place of $\delta^\mathrm{Pe}$ in Eq.~\ref{eq:pkernel}. Section~\ref{sec:result_eo} makes use of the ESC kernel for hydrodynamic spreading and interpolation, with the parameters $w=7$ and $\beta=7$ used for the cations, \textcolor{black}{and $w=4$, $\beta=8$ used for the anions.} This allows us to match the hydrodynamic radii of both sodium and chloride ions (in SPC water) on a common hydrodynamic grid. The four point Peskin kernel is used for electrostatic interactions.



\end{appendix}

\bibliography{FHDX}

\end{document}